\documentclass[12pt, a4paper]{article}
\usepackage[utf8]{inputenc}
\usepackage[english]{babel}
\usepackage[a4paper,top=2cm,bottom=2.5cm,left=2.5cm,right=2.5cm,marginparwidth=1.75cm]{geometry}
\usepackage{amsmath,amsfonts}
\usepackage{amssymb}
\usepackage{caption,adjustbox}
\usepackage{subcaption}
\usepackage{graphicx}
\usepackage{parskip}
\usepackage[colorlinks=true,linkcolor=Black,citecolor=Black,urlcolor=Blue]{hyperref}
\usepackage[usenames,dvipsnames]{xcolor}
\usepackage{physics, tensor, cancel}
\usepackage{tikz}
\usepackage[affil-sl]{authblk}

\usepackage{comment}
\usepackage{csquotes}
\usepackage{hyperref}
\usepackage[normalem]{ulem}
\usepackage{fancyhdr}
\usepackage{svg}
\usepackage[dvipsnames]{xcolor}
\usepackage{tcolorbox}
\usepackage{arydshln}

\tcbuselibrary{listings, theorems, skins, xparse, breakable}

\usepackage{wrapfig}

\usepackage[style=numeric-comp,sorting=none,maxbibnames=9]{biblatex}
\newcommand{\ellp}{\ell_\mathrm{P}}
\newcommand{\gnewton}{G_\mathrm{N}}
\newcommand{\sr}{r_\mathrm{S}}

\newcommand{\dho}{d_H}
\newcommand{\zh}{z_H}

\newcommand{\zhp}{z_{H,+}}
\newcommand{\zhm}{z_{H,-}}
\newcommand{\zhpm}{z_{H,\pm}}

\newcommand{\zbr}{z_{\text{BR},H}}

\newcommand{\dhp}{d_{H,+}}
\newcommand{\dhm}{d_{H,-}}

\newcommand{\zi}[1]{\mathfrak{p}_{#1}}

\newcommand{\fn}[1]{f^{(#1)}_{H}}
\newcommand{\hn}[1]{h^{(#1)}_{H}}

\newcommand{\fid}[1]{\Phi^{(#1)}_{H}}
\newcommand{\pid}[1]{\Psi^{(#1)}_{H}}

\newcommand{\ha}[1]{\widehat{a}_{#1}}

\newcommand{\rootk}{\varpi}

\usepackage{orcidlink}

\title{
\begin{flushright}{\vspace{-1.5cm}\small LYCEN 2023-02\\}\end{flushright}
\vspace{2.3cm}
\bf \Large Living on the Edge:\\[0.3cm] Quantum Black Hole Physics from the Event Horizon}
\author[1,3,5]{Manuel Del Piano\thanks{{\Large \orcidlink{0000-0003-4515-8787}} \href{mailto:manuel.delpiano-ssm@unina.it}{manuel.delpiano-ssm@unina.it}}}
\author[4]{Stefan Hohenegger\thanks{{\Large \orcidlink{0000-0001-6564-0795}} \href{mailto:s.hohenegger@ipnl.in2p3.fr}{s.hohenegger@ipnl.in2p3.fr}}} 
\author[1,2,3,5]{Francesco Sannino\thanks{{\Large \orcidlink{0000-0003-2361-5326}} \href{mailto:sannino@qtc.sdu.dk}{sannino@qtc.sdu.dk}}}

\affil[1]{\small Scuola Superiore Meridionale, Largo S. Marcellino, 10, 80138 Napoli NA, Italy}
\affil[2]{\small Dept. of Physics E. Pancini, Università di Napoli Federico II, via Cintia, 80126 Napoli, Italy}
\affil[3]{\small INFN sezione di Napoli, via Cintia, 80126 Napoli, Italy}
\affil[4]{\small Institut de Physique des 2 \textbf{}Infinis (IP2I), CNRS/IN2P3, UMR5822, 69622 Villeurbanne, France Universit\'e de Lyon, Universit\'e Claude Bernard Lyon 1, 69001 Lyon, France}
\affil[5]{\small Quantum  Theory Center QTC \& D-IAS, Southern Denmark Univ., Campusvej 55, 5230 Odense M, Denmark}
\date{}

\begin{document}

\maketitle
\begin{abstract}
\noindent
Quantum gravity theories predict deformations of black hole solutions relative to their classical counterparts. A model-independent approach was advocated in \cite{Binetti:2022xdi} that uses metric deformations parametrised in terms of physical quantities, such as the proper distance. While such a description manifestly preserves the invariance of the space-time under coordinate transformations, concrete computations are hard to tackle since the distance is defined in terms of the deformed metric itself. In this work, for spherically symmetric and static metrics, we provide a self-consistent framework allowing us to compute the distance function in close vicinity to the event horizon of a black hole. By assuming a minimal degree of regularity at the horizon, we provide explicit (series) expansions of the metric. This allows us to compute important thermodynamical quantities of the black hole, such as the Hawking temperature and entropy, for which we provide model-independent expressions, beyond a large mass expansion. Moreover, imposing for example the absence of curvature singularities at the event horizon leads to non-trivial consistency conditions for the metric deformations themselves, which we find to be violated by some models in the literature.

\end{abstract}

\newpage 
\tableofcontents

\section{Introduction}
Black holes are solutions of General Relativity (GR) with an event horizon, which potentially conceals a central singularity. While quantum corrections are expected to modify these solutions, our lack of a comprehensive theory of quantum gravity makes it challenging to provide precise details about these corrections. Over the years, numerous proposals for black hole deformations have emerged, drawing from both fundamental theories of gravity and effective approaches \cite{dymnikova1992vacuum, Bonanno:2000ep, Bjerrum-Bohr:2002fji, Gonzalez:2015upa, Nicolini:2019irw, Ruiz:2021qfp, Hayward_2006, Platania:2019kyx, Knorr:2022kqp, Binetti:2022xdi, Eichhorn:2022bgu, bardeen1968proceedings}. Notably, certain previous works \cite{Bonanno:2000ep, Binetti:2022xdi} have put forward deformations of the Schwarzschild space-time geometry, incorporating quantum corrections in a manner that allows for the formulation of universal statements.

Concretely, spherically symmetric and static space-time metrics are characterized by two functions $f$, $h$ of a radial coordinate \cite{Misner:1973prb, Carroll:2004st,penrose2005reality, hartle2003gravity}. We shall measure the latter in units of the Planck length $\ell_{\rm P}$ and denote it $z$. In general, assuming that $h \neq f$ allows for the description of a wide range of classical scenarios \cite{stephani_kramer_maccallum_hoenselaers_herlt_2003}. This includes the Tolman-Oppenheimer-Volkov space-time for compact objects and stellar environments \cite{Oppenheimer:1939ne, Boshkayev:2020kle, Boshkayev:2021chc, Kurmanov:2021uqv}.
A horizon in these geometries corresponds to a zero of the functions $f$ and $h$, which, for simplicity, in this work, we shall consider to be a simple zero. Asymptotically flat geometries are characterized by the fact that both $f$ and $h$ tend to $1$ for large values of $z$. 

Quantum corrections to the geometries mentioned above can be incorporated as deformations of the classical functions $f$ and $h$ and the precise form evidently depends on a concrete model of quantum gravity. Nevertheless, inspired by the \emph{renormalization group framework} \cite{ChenGoldenfeldOono1995,PhysRevE.49.4502,Barenblatt} it has been argued in \cite{Binetti:2022xdi,Bonanno:2000ep,DAlise:2023hls} that universal, model-independent statements about physical quantities in spherically symmetric and static quantum black holes can be made by demanding independence with respect to spurious scales. Notably, in order to preserve the invariance of the geometry under coordinate transformations (similar to those in GR), it has been advocated to write the deformation functions in terms of a physical quantity. While the concrete choice of the latter is to some degree ambiguous,\footnote{From the perspective of the renormalization group approach a different choice of this physical quantity (see \emph{e.g.} \cite{Borissova:2022jqj,Bonanno:2016dyv,Bonanno:2017zen,Held:2021vwd,Eichhorn:2022bgu}) corresponds to a different scheme. We shall elaborate on this connection in future work~\cite{InProgress}.} a natural choice in the context of static, spherically symmetric geometries is the proper distance from the center of the black hole. However, since the proper distance is defined in terms of the (deformed) metric function $f$, this prescription leads to an implicit definition of the quantum geometry. This issue has been addressed in previous works through different approximations: either by replacing the proper distance with a simpler function of $z$ (see \cite{Bonanno:2000ep}) or by assuming a very heavy black hole \cite{Binetti:2022xdi}. In the former case, the approximated distance generally no longer represents a physical quantity and therefore constitutes a conceptual departure from the above-mentioned logic. In the latter case, quantum corrections to physical quantities are suppressed by inverse powers of the mass. 

In this work, we provide a framework that allows us to compute the proper distance near the event horizon in a self-consistent fashion without the need for approximations. The framework assumes a certain degree of regularity of either the metric functions or the proper distance, such that they afford series expansions, at least up to some order. Furthermore, apart from the proper distance of the horizon from the center of the black hole $\dho$, the framework only requires information about the black hole exterior, which is encoded in the deformations of the metric functions. Concretely, within this setup, we find explicit solutions of the non-linear first-order differential equation that defines the proper distance in terms of the metric functions. These solutions completely determine the space-time geometry near the event horizon, which in turn allows us to compute the thermodynamical properties of the black hole, namely its Hawking temperature and entropy. Further assuming a dependence of $\dho$ on the mass of the black hole, we calculate mass expansions of the Hawking temperature, correcting previous results in \cite{Binetti:2022xdi}.

Furthermore, using this framework we find non-trivial conditions of the quantum-deformed black hole geometries. On the one hand, regularity of the first derivative of the metric functions is required to render the surface gravity well-defined, which in turn is required for the Hawking temperature to be well-defined \cite{Bardeen:1973gs}. However, this is not automatic but requires conditions on the metric deformations. On the other hand, the absence of curvature singularities at the event horizon also imposes non-trivial constraints: these can either be found by demanding finiteness of the second derivatives of the metric functions (which provides sufficient conditions) or by calculating series expansions of the Ricci and Kretschmann scalar close to (but outside of) the exterior event horizon. These generically exhibit divergent contributions which can only be removed if certain conditions for the metric deformations are met. We formulate these various conditions in the form of constraints on the original input parameters of the black hole geometry as mentioned before. Checking these conditions for certain examples in the literature, we find that they are not always respected. Indeed, the quantum black hole model proposed in \cite{Bonanno:2000ep} (which we shall refer to as the Bonanno-Reuter space-time) is based on a deformation of the metric functions that violate the above-mentioned conditions. This means, treating the Bonanno-Reuter space-time in a self-consistent fashion leads to an ill-defined Hawking temperature as well as a divergence of the Ricci scalar at the event horizon. While approximations to this space-time that were proposed in \cite{Bonanno:2000ep} do not suffer from such unphysical singularities, they should be interpreted as metric deformations different from the original ones, which comply with our consistency conditions. To further show the flexibility of our approach, we discuss as a different model a minimal solution to our conditions along with its physical properties.

Finally, in order to make closer contact with our previous work \cite{Binetti:2022xdi} we consider asymptotic expansions of the metric deformations in inverse powers of the proper distance. Assuming that the radius of convergence of these expansions is large enough to be still valid at the event horizon, we use them as input to the framework explained above. Concretely, we formulate the consistency conditions for the absence of an unphysical singularity in terms of the asymptotic expansion coefficients. Solving these conditions, we furthermore provide self-consistent expressions for the Hawking temperature and (upon assuming a mass-dependence of $\dho$) the entropy.

While our results are derived with a black hole geometry in mind, they can be generalized in a straightforward manner to spherically symmetric and static space-times. Therefore, we foresee the further impact of the framework presented here in the description of quantum effects in gravitational experiments and cosmology. Furthermore, while in this work we mainly have deformations of black holes in mind that are due to quantum effects, our approach is versatile enough to also describe other types of deformations. We, therefore, expect our work to be useful for studying space-time corrections in theories of modified gravity.

This paper is organised as follows: in Section~\ref{Sec2} we introduce our notation for deformations of the spherically symmetric and static Schwarzschild geometry. We derive non-trivial conditions for these deformations by imposing finiteness of the first and second derivative of the metric functions at the horizon: the first derivative is a necessary condition for the existence of the surface gravity and the second provides sufficient conditions for the absence of a singularity of the Ricci scalar at the horizon. In Section~\ref{Sect:SeriesExpansionApproach}, we develop a more general framework for computing the distance function for deformations of the Schwarzschild metric. Assuming the existence of a series expansion of the proper distance, we provide a recursive relation for all expansion coefficients. The consistency of this approach and the finiteness of the Ricci scalar at the horizon impose non-trivial conditions on the metric deformations. In Section~\ref{examples} we apply these conditions to concrete examples of black hole solutions. We show that the Bonanno-Reuter space-time does not abide by the conditions and indeed exhibits a singular behavior at the horizon. We also provide a novel model based on a minimalistic solution of the conditions derived previously. In Section~\ref{LDE} we consider asymptotic expansions of the metric deformations and (assuming that they can be extended all the way to the black hole horizon) show how to integrate them into the framework developed in previous Sections. Finally, Section~\ref{conclusions} contains our conclusions and an outlook for further applications. This work is complemented by 5 appendices:  Appendix~\ref{App:HigherOrderDerF} generalizes the approach of Section~\ref{Sec2} by deriving conditions imposed by assuming that an arbitrarily high order $N$ of derivatives of the metric function is finite at the horizon. Appendix~\ref{SolvingTheSystem} provides a minimal solution for the system of equations established in Appendix~\ref{App:HigherOrderDerF} and shows that this solution for $N\to \infty$ tends to the Schwarzschild space-time. Appendix~\ref{App:SerExpansions} contains several derivations of series identities that have been deemed too technical for the main body of the paper. Appendix~\ref{Sect:FurtherExamples} discusses further examples of deformed black hole metrics, namely the Hayward black hole and the Dymnikova space-time. Finally Appendix~\ref{App:InnerHorizon} gives a brief outline of how to generalize the conditions derived in Section~\ref{Sec2} to interior black hole horizons.


\section{Regular geometry close to the black hole horizon}\label{Sect:Conditions}
\label{Sec2}
Our starting point is the general form of a spherically symmetric and static space-time in four dimensions with Lorentzian signature
\begin{equation}\label{eq: metric}
    \dd{s}^2=g_{\mu \nu}\dd x^\mu \dd x^\nu=-h(r) \dd{t}^2+\frac{\dd{r}^2}{f(r)}+r^2 \dd{\theta}^2+r^2\sin^2\theta\dd{\phi}^2\ ,
\end{equation}
where the metric is given by
\begin{equation}
    g_{\mu \nu} = \mathrm{diag}(-h(r),f(r)^{-1},r^2,r^2 \sin^2{\theta} )\ ,
\end{equation}
with $h$ and $f$, \emph{a priori}, general functions of the radial coordinate $r$. The classical Schwarzschild space-time~\cite{Schwarzschild}, which is a solution of the Einstein equations in vacuum, is recovered for $h(r)=f(r)=(1-\sr/r)$, with $\sr=2 \gnewton M $ the Schwarzschild radius (and $\gnewton$ Newton's constant). For $r\geq \sr$, this metric describes the space-time outside of a central body of mass $M$.  In the following, we shall be interested in deformations of this metric, which specifically represent black holes and which are characterised by (particular) modifications of the metric functions $f$ and $h$.

To describe these modifications, we first simplify the notation by casting (\ref{eq: metric}) into a dimensionless form: similar to \cite{Binetti:2022xdi}, we write the radial coordinate (and the mass parameter $M$) in units of the Planck length $\ell_{\rm P}= 1/M_\mathrm{P}$ (with $M_\mathrm{P}$ the Planck mass), by defining:
\begin{equation}\label{eq: dimless z}
    z:=M_\mathrm{P}r=\frac{r}{\ell_\mathrm{P}} \qq{and} \chi:=\frac{M}{M_\mathrm{P}}\ .
    \end{equation} 
We shall further choose units such that Newton's constant is equal to $1$, $\emph{i.e.}$ $\gnewton\,M_\mathrm{P}^2=1$. In this notation, we shall parametrise deformations of the Schwarzschild geometry by writing the functions $f$ and $h$ as
\begin{equation}
    f(z) = 1 - \frac{2 \chi}{z}e^{\widetilde{\Phi}\left(z \right)}\, , \quad  h(z)=1-\frac{2 \chi}{z}e^{\widetilde{\Psi}\left(z \right)}\, ,
\end{equation}
where $\widetilde{\Phi}$ and $\widetilde{\Psi}$ encode corrections due to physical effects beyond GR, either classical or quantum in nature. In order to describe a black hole, we first require that the geometry is still asymptotically flat. Concretely, we assume that the geometry approaches the Schwarzschild metric (with mass parameter $\chi$) for very large distances from the origin
 \begin{equation}\label{asymptotic flatness}
 \lim_{z\to \infty}\widetilde{\Phi}(z) = 0 =\lim_{z\to \infty} \widetilde{\Psi}(z)\,. 
\end{equation}
Furthermore, in order for coordinate transformations of the undeformed space-time to be also realised in the deformed case, we demand that $\widetilde{\Psi}$ and $\widetilde{\Phi}$ are invariant quantities. This can be achieved by writing them as functions of a physical quantity, for which in \cite{Bonanno:2000ep,Binetti:2022xdi} the proper distance from the origin was proposed. The proper (radial) distance between two spatial points within the space-time (\ref{eq: metric}) is:
\begin{equation}
    d(z,z_0):=\int_{z_0}^z \frac{\dd{\tilde{z}}}{\sqrt{\abs{f(\tilde{z})}}}\, \quad {\rm and}  \quad d(z) = d(z,0) \ .
\label{propd}
\end{equation}
We thus replace in (\ref{modifiedfh})
\begin{equation}
        \widetilde{\Phi} (z),\, \widetilde{\Psi} (z) \quad \longrightarrow  \quad   \Phi (1/d(z)), \, \Psi (1/d(z))  \ ,
    \end{equation}
where $\Phi$ and $\Psi$ are functions of the inverse distance such that
\begin{equation}\label{modifiedfh}
    f(z) = 1 - \frac{2 \chi}{z}e^{\Phi\left(1/d(z) \right)}\, , \quad  h(z)=1-\frac{2 \chi}{z}e^{\Psi\left(1/d(z) \right)}\, ,
\end{equation}
Here we choose a dependence on $1/d$, such that asymptotical flatness~(\ref{asymptotic flatness}) amounts to the simple relation
\begin{align}
\Phi(0)= 0 =\Psi(0)\,.
\end{align}
Furthermore, in order to describe a black hole geometry, we require that the metric (\ref{eq: metric}) has (at least one) horizon, \emph{i.e.} we impose that the functions $f$ and $h$ have a zero at a coordinate $\zh\in(0,\infty)$
\begin{align}
f_{H}:=f(\zh)=0=h(\zh)=:h_H\,.\label{VanishingMetricFunctions}
\end{align}
In this paper, we shall assume that \eqref{VanishingMetricFunctions} are simple zeroes and that
\begin{align}
&f(z)>0\,,\qq{and} h(z)>0 \quad \forall z>\zh\,,
\end{align}
\emph{i.e.} that $\zh$ is the location of the outer horizon of the black hole. Furthermore, we introduce the notation
\begin{align}
&\dho:=d(\zh)\,,\qq{and} d(z)=\dho+\rho(z)\,,
\end{align}
where $\rho(z)=d(z,\zh)$ has the interpretation as the proper distance measured from the horizon of the black hole. With this notation, the condition (\ref{VanishingMetricFunctions}) can equally be written in the form
\begin{equation}
     \Phi \left(\frac{1}{\dho}\right) =\Psi \left(\frac{1}{\dho}\right) =  \log \frac{\zh}{2\chi} \ . 
\end{equation}
This ensures the vanishing of the norm of time-like Killing vector $(K^t)^\mu=\delta^{\mu 0}$ required for the existence of an event horizon  
\begin{equation}
    (K^t)_\mu (K^t)^\mu |_{z=\zh} = g_{00}(K^t)^0 (K^t)^0|_{z=\zh} =- h(\zh) = 0 \, . \label{Keq}
\end{equation}
We remark that in the remainder of this paper, unless otherwise specified, we shall consider the space-time outside of the horizon of the black hole, \emph{i.e.} we shall only consider the region $z\geq \zh$ (or equivalently $\rho\geq 0$). Furthermore, we consider $\dho$ as an additional input into the (exterior) black hole geometry, which is in fact the only information about the interior of the black hole that is required in the following.

The surface gravity and the Ricci scalar are fundamental quantities that must be well-defined at and near the horizon. We shall discuss the surface gravity in more detail in Section~\ref{Sect:ThermoGeneral}, while here we give the definition of the Ricci scalar as a geometric quantity, which directly follows from (\ref{eq: metric}): it is a scalar quantity which appears in the equation of motion of the gravitational field (\emph{i.e.} the Einstein equations). Therefore, its regularity (notably at the horizon) in the deformed metric ensures that in the deformed case no
additional singularities arise beyond the classical ones, which shall be a central point in the analysis of this paper. The Ricci scalar in terms of $f$ and $h$ is\footnote{To save writing factors of $\ellp$, we have defined a dimensionless form of $R$ (and $K$), \emph{i.e.} it is measured in units of $\ellp$. Furthermore, from here on out we use the notation $F^{(n)}$ to indicate the $n$-th derivative of a function $F$ with respect to its argument. For example, we define 
\begin{align}
&f^{(n)}(z):=\dv[n]{f(z)}{z}\,,&&    \text{and} &&\Phi^{(n)}\left( \frac{1}{x} \right) := \left.\dv[n]{\Phi(y)}{y}\right|_{y=\frac{1}{x}}\ . \label{Deffid}
\end{align}
The subscript $_H$ denotes the evaluation of a quantity at the horizon, which corresponds to taking $z =\zh$, $d = \dho$, or $\rho = 0$, \emph{e.g.}
\begin{align}
&\fn{n}:=f^{(n)}(z=\zh)\,,&&\text{and} &&\fid{n}:=\Phi^{(n)}(1/\dho)\,.
\end{align}

}
\begin{equation}\label{eq: ricci scalar h e f}
R = -\frac{\left(z f^{(1)}+4 f\right) h^{(1)}}{2 z h} + \frac{f \, h^{(2)}}{h}  -\frac{2\big(z f^{(1)}+f-1\big)}{z^2} +\frac{ f (h^{(1)})^2}{2 h^2}\, .
\end{equation}
We mainly focus on the finiteness of the Ricci scalar at the horizon but, in some examples, we shall also examine the behavior of the Kretschmann scalar, given by
\begin{equation}
    K=\tensor{R}{_\rho _\sigma _\mu _\nu}\tensor{R}{^\rho ^\sigma ^\mu ^\nu}\ ,\label{KretschDef}
\end{equation}
where $\tensor{R}{_\rho _\sigma _\mu _\nu}=\tensor{g}{_\rho _\lambda} \tensor{R}{^\lambda _\sigma _\mu _\nu}$ is the fully covariant Riemann tensor.


\subsection{Near horizon constraints}\label{nhc}
By examining the expression of the Ricci scalar \eqref{eq: ricci scalar h e f} and the Kretschmann scalar (\ref{KretschDef}), it is evident that derivatives of the functions $f$ and $h$ up to the second order (in $(z-\zh)$) are required. Hence, we start by assuming that in the vicinity of the black hole horizon, located at $\zh$, they can be expanded using a Taylor series up to the second order, given by
\begin{align}
    f(z) &= f^{(1)}_H (z-\zh) + \frac{f^{(2)}_H}{2}(z-\zh)^2  + \mathfrak{o}((z-\zh)^2)\, ,\label{fexp}\\
    h(z) &= h^{(1)}_H (z-\zh) + \frac{h^{(2)}_H}{2}(z-\zh)^2  + \mathfrak{o}((z-\zh)^2)\, .\label{hexp}
\end{align} 
We remark that similar expansions have been already considered in the literature, for example in \cite{Stelle2015, Sarkar:2007uz}. In the following, we shall assume $f^{(1)}_H>0$ and $h^{(1)}_H>0$ such that $\zh$ is a simple zero of both $f$ and $h$, as well as the position of the outer event horizon of the black hole space-time (\emph{i.e.} $f$ and $h$ have no further zeroes for $z>\zh$ and both change signs at $z=\zh$).

The expansions  (\ref{fexp}) and (\ref{hexp}) also afford the following form of the infinitesimal proper distance \eqref{propd} from the horizon 
\begin{equation}
    \dd{\rho} = \frac{\dd{z}}{\sqrt{f^{(1)}_H (z-\zh) + \frac{f^{(2)}_H}{2}(z-\zh)^2}}+\mathfrak{o}((z-\zh)^2)\, ,
\end{equation}
which yields explicitly 
\begin{equation}
\label{rho of z}
    \rho (z) = \frac{2 \sqrt{z-\zh}}{\sqrt{f^{(1)}_H }}-\frac{f^{(2)}_H (z-\zh)^{3/2}}{6 (f^{(1)}_H) ^{3/2}} +\mathfrak{o}((z-\zh)^{3/2})\ . 
\end{equation}
The latter can be locally inverted so that we can write $z(\rho)$, which up to the fourth order reads: 
\begin{equation}
    z(\rho) = \zh + \frac{f^{(1)}_H}{4}\rho^2 + \frac{f^{(1)}_H f^{(2)}_H}{96}\rho^4 + \mathfrak{o}(\rho^4)\, .\label{Expz}
\end{equation}
This allows us to rewrite the Taylor expansions for $f$ and $h$ in terms of the physical, and therefore coordinate-invariant, distance $\rho$ as follows\footnote{By abuse of notation we shall write in the following $f(\rho)$ instead of $f(z(\rho))$.}
\begin{align}
    f(\rho) &= \frac{(f^{(1)}_H)^2 }{4}\rho ^2 +\frac{(f^{(1)}_H)^2 f^{(2)}_H}{24} \rho ^4 + \mathfrak{o}(\rho^4)\ , \label{fTay}\\
    h(\rho) &= \frac{f^{(1)}_H h^{(1)}_H}{4}  \rho ^2 + \frac{f^{(1)}_H  (3 f^{(1)}_H h^{(2)}_H+f^{(2)}_H h^{(1)}_H)}{96}  \rho ^4 + \mathfrak{o}(\rho^4)\ .\label{hTay}
\end{align}
The derivatives of $f$ with respect to $z$ are consequently computed as (see (\ref{Deffid}) for the notation)
\begin{equation}\label{fPhiDer1}
    f^{(1)}(z) = \dv{f}{z} = \dv{f(\rho(z))}{\rho}\dv{\rho}{z} = \frac{1}{\sqrt{f(\rho)}}\dv{f(\rho)}{\rho}\, ,
\end{equation}
and similarly for $h^{(1)}$. In terms of the functions $\Phi$ and $\Psi$ in (\ref{modifiedfh}) we therefore find 
 \begin{align}
     f^{(1)}(z)=(1-f)\left(\frac{1}{z}+\frac{\Phi^{(1)}\left( \frac{1}{\dho+\rho}\right)}{(\dho+\rho)^2 \sqrt{f}} \right) \ ,\qq{and} h^{(1)}(z)=(1-h)\left(\frac{1}{z}+\frac{\Psi^{(1)}\left( \frac{1}{\dho+\rho}\right)}{(\dho+\rho)^2 \sqrt{f}} \right) \ .\label{fprimePhi}
 \end{align}
 It is clear from the above that the first derivative shows a divergence at the horizon due to the second term in the parenthesis $\propto f^{-1/2}$. Using \eqref{Expz}, \eqref{fTay}, and \eqref{hTay} we can provide a series expansion around $\rho = 0$, concretely for $f^{(1)}$:
 \begin{align}
     f^{(1)} &= \frac{2 \fid{1}}{\dho ^2 \fn{1}}\frac{1}{\rho} + \frac{1}{\zh} - \frac{2\left(2\dho \fid{1} + \fid{2}\right)}{\dho^4 \fn{1}} + \mathfrak{o}(\rho^0)\,,\label{fderFirst}
 \end{align}
and similarly for $h^{(1)}$. Therefore, we conclude that consistency with the first order in~\eqref{fexp} and \eqref{hexp} requires
 \begin{equation}\label{phip}
    \Phi^{(1)}_H = 0 \, , \, \quad {\rm and}  \quad\Psi^{(1)}_H = 0\, ,
\end{equation}
which removes the  singularity in~\eqref{fprimePhi}. 
With these conditions, we get the expressions of the first derivatives of the metric functions at the event horizon\footnote{We remark that $\fn{1}=\frac{1-\rootk}{2\zh}$ is also a solution compatible with (\ref{fderFirst}). In the following, we shall use the result (\ref{f1h h1h}), which leads to the classical Schwarzschild metric for $\Phi^{(2)}_H=0$. }
\begin{align}\label{f1h h1h}
 \fn{1} = \frac{1+\rootk}{2 \zh}\qq{and}  h^{(1)}_H = \frac{1}{\zh}-\frac{2 \Psi^{(2)}_H}{\dho^4 \fn{1}}\, ,
\end{align} 
where for later use we have introduced the shorthand notation
\begin{align}
\rootk=\sqrt{1-\frac{8\zh^2 \fid{2}}{\dho^4}}\,.\label{DefRootK}
\end{align}
Reality of $\fn{1}$ requires that $\rootk\in\mathbb{R}$. Furthermore, in order for $\zh$ to be the position of (a simple) outer event horizon, both $\fn{1}>0$ and $h^{(1)}_{H}>0$. These conditions together impose upper bounds for the derivatives $\fid{2}$ and $\pid{2}$ : 
\begin{align}
&\Phi^{(2)}_H \leq \frac{\dho^4}{8\zh^2}\,,\qq{and} \pid{2}<\frac{\dho^4(1+\rootk)}{4\zh^2}\,.\label{Bounds}
\end{align}
Moving to the second order of the Taylor polynomials in~\eqref{fexp} and~\eqref{hexp}, we find for the second derivatives
\begin{align}
    f^{(2)} &= -f^{(1)}\left(\frac{1}{z}+\frac{\Phi^{(1)}\left( \frac{1}{\dho+\rho}\right)}{(\dho+\rho)^2 \sqrt{f}} \right)+\nonumber \\
    &-(1-f) \left(\frac{1}{z^2} + \frac{2\Phi^{(1)}\left( \frac{1}{\dho+\rho}\right)}{(\dho+\rho)^3 f} + \frac{f^{(1)}\Phi^{(1)}\left( \frac{1}{\dho+\rho}\right)}{2 (\dho+\rho)^2 f^{3/2}} +\frac{\Phi^{(2)}\left( \frac{1}{\dho+\rho}\right)}{(\dho+\rho)^4 f^{3/2}}\right) ,\nonumber\\
    h^{(2)} &= -h^{(1)}\left(\frac{1}{z}+\frac{\Psi^{(1)}\left( \frac{1}{\dho+\rho}\right)}{(\dho+\rho)^2 \sqrt{f}} \right)+\nonumber \\
    &-(1-h) \left(\frac{1}{z^2} + \frac{2\Psi^{(1)}\left( \frac{1}{\dho+\rho}\right)}{(\dho+\rho)^3 f} + \frac{f^{(1)}\Psi^{(1)}\left( \frac{1}{\dho+\rho}\right)}{2 (\dho+\rho)^2 f^{3/2}} +\frac{\Psi^{(2)}\left( \frac{1}{\dho+\rho}\right)}{(\dho+\rho)^4 f^{3/2}}\right),
\end{align}
which we can equally expand around $\rho = 0$, similar to eq.~\eqref{fprimePhi}. By utilising \eqref{phip} and the expressions for $\fn{1}$ and $\hn{1}$ given in equation \eqref{f1h h1h}, we can eliminate the divergent terms by imposing that
\begin{equation}
    \Phi^{(3)}_H = - 6 \dho \Phi^{(2)}_H  \qq{and} \Psi^{(3)}_H = - 6 \dho \Psi^{(2)}_H\ .
\end{equation}
As explained in appendix~\ref{App:AllHigherOrders}, demanding the finiteness of even higher order derivatives at the horizon (\emph{i.e.} beyond second order), one can iterate the above procedure of removing divergent terms in the Taylor expansion, leading to the general result (\ref{XiHigherRel}). 

Summarising the result for imposing finiteness of the first and second derivatives of $f$ and $h$ (as in \eqref{fexp} and~\eqref{hexp}), as well as the reality and positiveness of the first derivatives, we have obtained the following general constraints and upper bounds: 
\begin{tcolorbox}[ams equation,colback=black!10!white,colframe=black!95!green]
\parbox{14cm}{${}$\\[-40pt]\begin{align}&\Phi_H = \Psi_H = \log\frac{\zh}{2\chi} \ , \quad \Phi^{(1)}_H = \Psi^{(1)}_H = 0 \ , \quad   \Phi^{(3)}_H = - 6 \dho \Phi^{(2)}_H\ , \quad \Psi^{(3)}_H = - 6 \dho \Psi^{(2)}_H\ ,
\nonumber\\
&\hspace{4cm}\fid{2}\leq \frac{\dho^4}{8\zh^2}\,,\hspace{1.5cm} \Psi^{(2)}_H<\frac{\dho^4(1+\rootk)}{4\zh^2}\,.\label{constraints}\end{align}\nonumber ${}$\\[-30pt]}
\end{tcolorbox}
By imposing these conditions, we ensure that there are no curvature singularities at the horizon: indeed, with (\ref{eq: ricci scalar h e f}) (and (\ref{KretschDef})) it can be verified that the Ricci and Kretschmann scalars attain finite values, which depend on $\fid{2}$, $\fid{4}$, $\pid{2}$, and $\pid{4}$.\footnote{Due to their complexity, we refrain from presenting the complete expressions of $R$ and $K$.}
 
However, to provide a more comprehensive and generalized framework, we delve into a broader approach that encompasses these conditions in Section~\ref{SectMetricFunctCurv}.

\subsection{Impact on the thermodynamics}\label{Sect:ThermoGeneral}
One of the main applications of the general constraints discussed above is black hole thermodynamics. To determine the temperature of a black hole we introduce 
the \emph{surface gravity} which is expressed via the time-like Killing vector $(K^t)^{\mu}$ introduced earlier (see eq.~(\ref{Keq})) and reads \cite{Sarkar:2007uz,Wald:1984rg,Bogoliubov1958,Hawking1975,birrell_davies_1982}
\begin{equation}\label{surf grav def}
    \kappa^2=\left.-\frac{\ellp^2}{2}\nabla_\mu (K^t)_\nu \nabla^\mu (K^t)^\nu \right|_{z=z_H} = \left.\frac{1}{4}\frac{f\cdot (h^{(1)})^2}{h}\right|_{z=z_H}\, . 
\end{equation}
Using the constraints in \eqref{constraints} the Hawking temperature \cite{Hawking1975,Kay:1988mu} is then given by\footnote{In order to avoid factors of $\ellp$ in the following, we have defined dimensionless versions of both the surface gravity (\ref{surf grav def}) and the Hawking temperature (\ref{Tempkappa}), which are measured in units of $\ellp$.}
\begin{tcolorbox}[ams align,colback=black!10!white,colframe=black!95!green]\label{Tempkappa}
     T_\mathrm{H}=\frac{\kappa}{2\pi} = \frac{1}{4 \pi}\sqrt{f^{(1)}_H h^{(1)}_H} = \frac{1}{4\pi}\sqrt{\frac{1+\rootk}{2\zh^2}-\frac{2\Psi^{(2)}_H}{\dho^4}}\, ,&&\text{with} &&\rootk=\sqrt{1-\frac{8\zh^2 \fid{2}}{\dho^4}}\,.
\end{tcolorbox}
Notice that due to the upper bound on $\Psi_H^{(2)}$, the Hawking temperature is strictly positive $T_{\text{H}}>0$ for $\dho>0$ (and thus $\zh>0$). Indeed, $T_{\text{H}}=0$ would require $\fn{1}=0$ and/or $h_H^{(1)}$. With (\ref{fexp}) and (\ref{hexp}), this would, however, imply that $f$ and/or $h$ would have a double zero at the horizon, which is thus not compatible with our initial assumption in this approach.

For the expression of the entropy, we use the first law of thermodynamics
\begin{equation}
   \dd \chi = T_\mathrm{H} \dd S\, .
\end{equation}
such that the entropy can be written as
\begin{equation}\label{EntropyDef}
    S = \int \frac{\dd \chi}{T_\mathrm{H}(\chi)}\, .
\end{equation}
To perform the integration over the black hole mass $\chi$, one has to provide the explicit dependence of $\dho$, $\zh$, $\Psi^{(2)}_H$ and $\Phi^{(2)}_H$ on $\chi$ itself.  

\section{Series expansions of the distance function}\label{Sect:SeriesExpansionApproach}
In the previous Section, we have provided the conditions (\ref{constraints}) for the functions $\Phi,\Psi$ appearing in eq.~(\ref{modifiedfh}) that guarantee finite first and second derivatives of the metric functions at the event horizon. These in turn are sufficient conditions that also the Ricci and Kretschmann scalar are finite. In this Section, we shall recover more general (but compatible) results by using a different approach, namely by solving (\ref{propd}) assuming that a solution in the form of a series expansion exists.
\subsection{Solving the distance function}\label{Sect:DistanceFunctionExpansion}
\subsubsection{Series expansion of the radial coordinate}\label{Sect:SeriesExpansion}
In eq.~(\ref{rho of z}) we have already given a limited series expansion of the distance to the BH horizon $\rho$ as a function of $z-\zh$, the inversion of which is given in (\ref{Expz}). In this Section, we shall provide a general form of these expansions under the assumption that $z$ can be written as a (convergent) series in $\rho$ for $\rho\geq 0$. Indeed, we shall start from a general (integer) series of the form\footnote{Other than in the previous Section, we presently do not make any assumptions on (derivatives of) $f$ or $h$.}
\begin{align}
&z(\rho)=\zh+\sum_{n=1}^\infty a_n\,\rho^n\, &&\text{with} &&a_{n}\in\mathbb{R}\hspace{0.5cm}\forall n\in\mathbb{N}\,,\label{SeriesFormaInt}
\end{align}
which we assume to have an interval of convergence $\rho\in[0,\rho_A)$ for some $\rho_A>0$. We shall determine the series coefficients $a_{n}$ recursively by solving the differential equation
\begin{align}
&\dv{\rho}{z}=\left(1-\frac{2\chi}{z}\,e^{\Phi\left(\frac{1}{\dho+\rho}\right)}\right)^{-1/2}\,,&&\forall z\geq \zh\,.\label{SeriesForm}
\end{align}
However, we shall consider $z$ as a function of $\rho$, \emph{i.e.} we consider (\ref{SeriesForm}) in the form 
\begin{align}
&z\left(1-\left(\dv{z}{\rho}\right)^2\right)=2\chi\,e^{\Phi\left(\frac{1}{\dho+\rho}\right)}\,,&&\forall \rho\geq 0\,.\label{DiffEqSeries}
\end{align}
We next assume that $2\chi\,e^{\Phi}$ affords a series expansion in powers of $\rho$
\begin{align}
&2\chi\,e^{\Phi\left(\frac{1}{\dho+\rho}\right)}=\sum_{n=0}^\infty\,\xi_n\,\rho^n\,,&&\text{with} &&\begin{array}{l}\xi_0=\zh\,,\\\xi_n\in\mathbb{R}\hspace{0.5cm}\forall n\in\mathbb{N}\end{array}\,,\label{XiCoefs}
\end{align}
which has an interval of convergence $I\supseteq [0,\rho_A)$. Concretely, the coefficients $\xi_n$ can be related to the $\fid{n}$ as introduced in (\ref{Deffid}): starting from the expansion
\begin{align}
&\Phi\left(\frac{1}{\dho+\rho}\right)=\sum_{n=0}^\infty\kappa_n\,\rho^n\,,\label{SerPhiKappa}
\end{align}
with the coefficients
\begin{align}
\kappa_n=\frac{(-1)^n}{n!}\,\left[\frac{\fid{n}}{\dho^{2n}}+\sum_{k=1}^n\frac{\left(\prod_{i=1}^{k-1}(n-i)\right)\left(\prod_{j=0}^k(n-j)\right)}{k!\,\dho^{2n-k}}\,\fid{n-k}\right]\,,\label{KappaRelPhi}
\end{align}
we have verified the following relation up to order $n=8$
\begin{align}
\xi_n=2\chi\,e^{\Phi_H}\sum_{k=0}^n\sum_{\{u_1,\ldots,u_n\}}\frac{\kappa_1^{n-k}\,\prod_{i=2}^n\kappa_i^{u_i}}{(n-k)!\prod_{j=2}^n (u_j!)}\,,&&\text{with} &&u_j\in\mathbb{N}\cup\{0\}\text{ such that }\sum_{j=2}^n j u_j=k\,.\label{XiRelPhi}
\end{align}
For example, we find for the first three coefficients
\begin{align}
\xi_0=2\chi\,e^{\Phi_H}=\zh\,,&&\xi_1=-\frac{2\chi\,e^{\Phi_H}}{\dho^2}\,\fid{1}\,,&&\xi_2=\frac{\chi\,e^{\Phi_H}}{\dho^4}\,\left(2\dho\,\fid{1}+\left(\fid{1}\right)^2+\fid{2}\right)\,.\label{XiCoefsExplicit}
\end{align}
In the following, we shall consider the coefficients $\xi_n$ (or equivalently the derivatives $\fid{n}$ of $\Phi$ at the horizon) along with $\dho$ as fixed and as input for the equation (\ref{DiffEqSeries}). We can then expand the left-hand side of the latter into a series expansion in $\rho$. To this end, we consider
\begin{align}
\left(\dv{z}{\rho}\right)^2&=\sum_{n,m=1}^\infty n\,m\,a_{n}\,a_{m}\,\rho^{m+n-2}=a_1^2+\sum_{n=1}^\infty\left(\sum_{m=1}^{n+1}(n-m+2)\,m\,a_{m}\,a_{n-m+2}\right)\,\rho^{n}\,.\nonumber
\end{align}
Inserting this expansion into the left-hand side of (\ref{DiffEqSeries}) we find 
\allowdisplaybreaks{
\begin{align}
&z\left(1-\left(\dv{z}{\rho}\right)^2\right)
=\zh\,(1-a_1^2)+\sum_{n=1}^\infty\left[(1-a_1^2)\,a_n-\zh\,\sum_{m=1}^{n+1}(n-m+2)\,m\,a_m\,a_{n-m+2}\right]\,\rho^n\nonumber\\
&\hspace{5.5cm}-\sum_{p=1}^\infty\left[\sum_{n=1}^{p-1}\sum_{m=1}^{n+1}(n-m+2)\,m\,a_{p-n}\,a_{m}\,a_{n-m+2}\right]\,\rho^p\,,
\end{align}}
such that equation (\ref{DiffEqSeries}) becomes
\begin{align}
\sum_{p=0}^\infty \xi_p\rho^p=\zh\,(1-a_1^2)&+\sum_{p=1}^\infty\bigg[(1-a_1^2)\,a_p-\zh\,\sum_{m=1}^{p+1}(p-m+2)\,m\,a_m\,a_{p-m+2}\nonumber\\
&\hspace{1.5cm}-\sum_{n=1}^{p-1}\sum_{m=1}^{n+1}(n-m+2)\,m\,a_{p-n}\,a_m\,a_{n-m+2}\bigg]\rho^p\,.\label{SeriesEquation}
\end{align}
Identifying the series coefficients order by order for $p\in\{0,1,2\}$ leads to
\begin{align}
&\xi_0=\zh=\zh(1-a_1^2)\,,\nonumber\\
&\xi_1=(1-a_1^2)a_1-4\,\zh\,a_1\,a_2\,,\nonumber\\
&\xi_2=(1-a_1^2)\,a_2-\left[\zh(6\,a_1\,a_3+4\,a_2^2)+4\,a_1^2\,a_2\right]\,.\label{InitialConditionsGenericExpansion}
\end{align}
The first of these equations requires $a_1=0$, which imposes the condition $\xi_1=0$.\footnote{We remark in passing that a series expansion for $z$ to match the case $\xi_1\neq 0$ requires half-integer powers in eq.~(\ref{SeriesFormaInt}), as is discussed in Appendix~\ref{App:SerNonZeroXi}.
We shall consider such a case in more detail in the context of a specific example in Appendix~\ref{Sect:BRDistanceComputations}. As we shall see, however, $\xi_1\neq 0$ (and thus $\fid{1}\neq 0$) leads in general to a curvature singularity at the horizon.} Using (\ref{XiCoefsExplicit}), this requires $\fid{1}=0$, which is in fact the second relation in (\ref{constraints}). The last equation in (\ref{InitialConditionsGenericExpansion}) then becomes
\begin{align}
\xi_2=a_2(1-4\,\zh\,a_2)\,,\label{a2Xi2}
\end{align}
which has solution 
\begin{align}
&a_2=\frac{1\pm\rootk}{8\,\zh} \qq{with} \rootk=\sqrt{1-16\zh\xi_2}\,.\label{Coefs2Extract}
\end{align}
For $\xi_1=0$ (and thus $\fid{1}=0$), $\rootk$ is identical to the definition in (\ref{DefRootK}), since in this case $\xi_2=\frac{\zh \fid{2}}{2\dho^4}$, which follows from (\ref{XiCoefsExplicit}). The solution (\ref{Coefs2Extract}) is in agreement with (\ref{Expz}) using (\ref{f1h h1h}).\footnote{Indeed, in (\ref{f1h h1h}) the solution with the $+$ sign has been chosen to recover the classical result in the limit $\fid{2}\to 0$. In the following, we shall make the same choice.} Similar to (\ref{Bounds}), the requirement for $\rootk$ to be real (such that $a_2\in\mathbb{R}$) imposes an upper bound on $\xi_2\leq1/(16\zh)$.

Equating the remaining powers of $\rho^p$ (for $p>2$) in (\ref{SeriesEquation}) then becomes a recursive equation, which allows to express $a_{p}$ in terms of coefficients $a_{p'}$ with $p'<p$ (and $\xi_p$):
\begin{align}
a_p=\frac{1}{1-4\,\zh\,p\,a_2}\bigg[\xi_p&+\zh\,\sum_{n=3}^{p-1}(p-n+2)\,n\,a_n\,a_{p-n+2}\nonumber\\
&+\sum_{n=2}^{p-2}\sum_{m=2}^n(n-m+2)\,m\,a_{p-n}\,a_m\,a_{n-m+2}\bigg]\,,\hspace{1cm}\forall p\geq 3\,.\label{RecursionP}
\end{align}
The equations (\ref{RecursionP}) allow us to compute the coefficients $a_{p}$ explicitly up to arbitrary order as functions of the $\xi_{n}$. 
\subsubsection{Series expansion of the distance and metric function}
In order to expand the metric function $f$ in a series in $z-\zh$ (as in eq.~(\ref{fexp})), we first assume an expansion of $\rho$ in (half-integer) powers of $z-\zh$
\begin{align}
\rho=\sum_{n=1}^\infty b_n\,(z-\zh)^{n/2}\,,\label{SeriesRho}
\end{align}
The coefficients $b_n$ can be computed from (\ref{RecursionP}) to arbitrarily high order using series reversion. Indeed, generalizing a result of Whittaker \cite{Whittaker} for the reversion of integer series with $a_1\neq0$ to the current case, we have verified up to order $n=8$
\begin{align}\label{Coeffbina}
&b_1=\frac{1}{a_2^{1/2}}\,,&&b_2=-\frac{a_3}{2a_2^2}\,,&&b_n=\frac{(-1)^{n-1}}{2^{n-1}n!\,a_2^{n/2}}\,\text{det}(\mathcal{M}_n)\hspace{0.5cm}\forall n\geq 3\,,
\end{align}
where $\mathcal{M}_n$ is the following $(n-1)\times (n-1)$ matrix
\begin{align}\scalebox{0.6}{\parbox{12cm}{\begin{align}
\mathcal{M}_n=\left(\begin{array}{c;{2pt/2pt}c;{2pt/2pt}c;{2pt/2pt}c;{2pt/2pt}c;{2pt/2pt}c;{2pt/2pt}c;{2pt/2pt}c;{2pt/2pt}c;{2pt/2pt}c} n\frac{a_3}{a_2} & 1 & 0 & 0 & 0& 0 & 0 & 0 & \cdots &0 \\[4pt]\hdashline[2pt/2pt]
& & & & & & & & &\\[-12pt]
4n\frac{a_4}{a_2} & (n+2)\frac{a_3}{a_2} & 2 & 0 & 0 & 0 &  0 & 0 & \cdots & 0 \\[4pt]\hdashline[2pt/2pt]
& & & & & & & & &\\[-12pt]
12n\frac{a_5}{a_2} & (4n+4)\frac{a_4}{a_2} & (n+4)\,\frac{a_3}{a_2} & 3 & 0 & 0 & 0 & 0 & \cdots & 0 \\[4pt]\hdashline[2pt/2pt]
& & & & & & & & &\\[-12pt]
32n\frac{a_6}{a_2} & (12n+8)\frac{a_5}{a_2} & (4n+8)\,\frac{a_4}{a_2} & (n+6)\frac{a_3}{a_2} & 4 & 0 & 0 & 0 &\cdots & 0 \\[4pt]\hdashline[2pt/2pt]
& & & & & & & & &\\[-12pt]
80n\frac{a_7}{a_2} & (32n+16)\frac{a_6}{a_2} & (12n+16)\,\frac{a_5}{a_2} & (4n+12)\frac{a_4}{a_2} & (n+8)\frac{a_3}{a_2} & 5 & 0 & 0 & \cdots & 0 \\[4pt]\hdashline[2pt/2pt]
& & & & & & & & &\\[-12pt]
\vdots & \vdots & \vdots & \vdots & \vdots & \ddots & \ddots & \cdots & \cdots & \vdots \\[4pt]\hdashline[2pt/2pt]
& & & & & & & & &\\[-12pt]
2^{k-1}kn\frac{a_{k+2}}{a_2} & 2^{k-2}((k-1)n+2)\frac{a_{k+1}}{a_2} & 2^{k-3}((k-2)n+4)\frac{a_k}{a_2} & \cdots & \cdots & \cdots & (n+2^{k-1})\frac{a_3}{a_2} & k& \cdots & 0 \\[4pt]\hdashline[2pt/2pt]
& & & & & & & & &\\[-12pt]
\vdots & \vdots & \vdots & \vdots & \vdots & \vdots & \vdots & \ddots & \ddots & \vdots \\[4pt]\hdashline[2pt/2pt]
& & & & & & & & &\\[-12pt]
2^{n-3}n(n-2)\frac{a_{n}}{a_2} & \vdots & \vdots & \vdots & \vdots & \vdots & \vdots & \ddots & \ddots & n-2 \\[4pt]\hdashline[2pt/2pt]
& & & & & & & & &\\[-12pt]
2^{n-2}n(n-1)\frac{a_{n+1}}{a_2} & 2^{n-3}(n(n-2)+2)\frac{a_n}{a_2} & \cdots & \cdots & \cdots & \cdots & \cdots & \cdots & \cdots & (n+2^{n-2})\frac{a_3}{a_2}
 \end{array}\right)\nonumber
 \end{align}
 }}\nonumber
\end{align}
Concretely, we find for the first few coefficients with $n\geq 3$
\begin{align}
& b_3=\frac{5a_3^2-4a_2 a_4}{8a_2^{7/2}}\,,\hspace{1cm}b_4=\frac{3 a_2 a_3 a_4-2 a_3^3-a_2^2 a_5}{2a_2^{5}}\,,\nonumber\\
&b_5=\frac{231 a_3^4-504 a_2 a_3^2 a_4+224 a_2^2 a_3 a_5+112 a_2^2 a_4^2-64 a_2^3 a_6}{128 a_2^{13/2}}\,.
\end{align} 
\subsection{Metric functions and curvature}\label{SectMetricFunctCurv}
In the previous Subsection, we solved the differential equation (\ref{SeriesFormaInt}), assuming an integer series expansion of the radial coordinate as a function of the distance (from the event horizon) $\rho$. In this Subsection, we use this result to compute the Ricci scalar. 
\subsubsection{Metric functions}
As a first step, similar to (\ref{XiCoefs}), we also introduce a series expansion for the deformation of the metric function $h$
\begin{align}
&2\chi\,e^{\Psi\left(\frac{1}{\dho+\rho}\right)}=\sum_{n=0}^\infty\,\theta_n\,\rho^n\,,&&\text{with} &&\begin{array}{l}\theta_0=\zh\,,\\\theta_n\in\mathbb{R}\hspace{0.5cm}\forall n\in\mathbb{N}\end{array}\,.\label{ThetaCoefs}
\end{align}
The coefficients $\theta_n$ can be related to the derivatives $\pid{n}$ in exactly the same way as the coefficients $\xi_n$ are related to the $\fid{n}$ in eq.~(\ref{XiRelPhi}). For the moment, we shall keep $\theta_1$ generic and not mimic the constraint $\xi_1=0$, which was required for consistency of the series expansion (\ref{SeriesFormaInt}) with (\ref{XiCoefs}).

Using (\ref{XiCoefs}), (\ref{ThetaCoefs}), as well as the expansion $1/z$ in (\ref{SeriesFormaInvZ}) (with the coefficients $\zi{m}$ given recursively in eq.~(\ref{1zRecursive})), we can express the metric functions as power series in $\rho$
\begin{align}
&f=1-\sum_{k=0}^\infty\rho^k\sum_{n=0}^k\,\zi{n}\,\xi_{k-n}\,,&&\text{and} &&h=1-\sum_{k=0}^\infty\rho^k\sum_{n=0}^k\,\zi{n}\,\theta_{k-n}\,,&&\forall \rho\geq 0\,,\label{fhSerRho}
\end{align}
which reads explicitly to leading orders 
\begin{align}
f&=\frac{a_2-\xi_2}{\zh}\,\rho^2+\frac{a_3-\xi_3}{\zh}\,\rho^3+\mathcal{O}(\rho^4)\,,\label{frho}\\
h&=-\frac{\theta_1}{\zh}\,\rho+\frac{a_2-\theta_2}{\zh}\,\rho^2+\frac{(a_3-\theta_3)\zh+a_2\theta_1}{\zh^2}\,\rho^3+\mathcal{O}(\rho^4)\,.\label{hrho}
\end{align}
We can equivalently write these as expansions in $(z-\zh)$
\begin{align}
f&=\frac{a_2-\xi_2}{a_2\,\zh}\,(z-\zh)+\frac{a_3\xi_2-a_2\xi_3}{(a_2)^{5/2}\zh}\,(z-\zh)^{3/2}+\mathcal{O}((z-\zh)^2)\,,\label{fzExpansion}\\
h&=-\frac{\theta_1}{\sqrt{a}_2\zh}\,(z-\zh)^{1/2}+\frac{2a_2^2+a_3\theta_1-2a_2\theta_2}{2a_2^2\,\zh}\,(z-\zh)+\mathcal{O}((z-\zh)^{3/2})\,,\label{hzExpansion}
\end{align}
which shows certain similarities with the results of Section~\ref{nhc}, but also some differences: on the one hand, comparing the expansion for the function $f$ with (\ref{fTay}), since 
\begin{align}
\frac{a_2-\xi_2}{a_2\,\zh}=4\,a_2=\frac{1+\rootk}{2\,\zh}=f^{(1)}_H\,,
\end{align}
(see (\ref{f1h h1h})) we find agreement with the leading term in (\ref{fzExpansion}). On the other hand, (\ref{fzExpansion}) contains a term of order $(z-\zh)^{3/2}$ (and (\ref{frho}) a term of order $\rho^3$), which is absent in (\ref{fexp}) (and (\ref{fTay}) respectively). Indeed (\ref{fzExpansion}) is not an integer series expansion and therefore more general than (\ref{fexp}). Similarly, (since we have not imposed $\theta_1= 0$), the function $h$ even starts from a term $(z-\zh)^{1/2}$ in (\ref{hzExpansion}) (order $\rho$ in (\ref{hrho})), which is absent in (\ref{hexp}) (and (\ref{hTay}) respectively). Our result (\ref{fhSerRho}) is therefore more general than (\ref{fTay}) and (\ref{hTay}). We remark, however, that for generic values of $\xi_3$ and $\theta_1$ we have
\begin{align}
&\lim_{z\to \zh}f^{(2)}(z)\to\infty\,,&&\text{and} &&\lim_{z\to \zh}h^{(1)}(z)\to\infty\,.\label{SingDers}
\end{align}
Indeed, in contrast to Section~\ref{nhc} finiteness of the derivatives of the metric functions at the horizon was not our initial assumption and we shall impose in the next Subsubsection absence of a physical curvature singularity instead as a more general condition. Before doing so, however, we comment that we can deduce the conditions for the absence of the singularities (\ref{SingDers}) from (\ref{fzExpansion}) and (\ref{hzExpansion}) in a straightforward manner:
\begin{align}
&\theta_1=0\,,&&\text{and} &&a_3\xi_2=a_2\xi_3\,,&&\text{and} &&a_3\theta_2=a_2\theta_3\,.
\end{align}
Using (\ref{RecursionP}), the second of these relations implies 
\begin{align}
\frac{1-8\zh\xi_2+\rootk}{2\zh(1+3\rootk)}\,\xi_3=0\,,
\end{align}
which has as only solution $\xi_3=0$, which further implies $\theta_3=0$. Notice that the conditions 
\begin{align}
\xi_1=\xi_3=\theta_1=\theta_3=0\,,\label{ConditionsfhRegular}
\end{align}
which guarantee existence of the first and second derivative of $f$ and $h$ for $z=\zh$ (which are necessary for the expansions (\ref{fexp}) and (\ref{hexp})) are precisely the same as (\ref{constraints}). In this case, the positivity of the first derivative of $h$ leads to an upper bound on $\theta_2<\frac{1+\rootk}{8\zh}$, which agrees with the second relation in (\ref{Bounds}). The upper bound $\xi_2\leq 1/(16 \zh)$ was already obtained previously to guarantee reality of $\rootk$ in (\ref{Coefs2Extract}).

\subsubsection{Ricci scalar and Hawking temperature}
We next consider the Ricci scalar, however, for simplicity, we shall work out its series expansions only to leading order. Moreover, we shall start out by only assuming $\xi_1=0$ (which is required for the consistent expansion (\ref{SeriesFormaInt})), but we shall not assume the remaining conditions in (\ref{ConditionsfhRegular}). We notably first consider $\theta_1\neq 0$. Inserting (\ref{fzExpansion}) and (\ref{hzExpansion}) into (\ref{eq: ricci scalar h e f}), we obtain the following series expansion (for $z\geq \zh$)
\begin{align}
R=\frac{a_2-\xi_2}{8a_2\zh}\,(z-\zh)^{-1}+\frac{(2a_2^2+a_3\theta_1-2a_2\theta_2)(a_2-\xi_2)}{8a_2^{5/2}\theta_1 \zh}\,(z-\zh)^{-1/2}+\mathcal{O}((z-\zh)^0)\,.\label{RicciSeriesExpansionThetaneq}
\end{align}
Using (\ref{RecursionP}) the coefficient of the leading term becomes
\begin{align}
\frac{a_2-\xi_2}{8a_2\zh}=\frac{1-8\zh\xi_2+\rootk}{8\zh(1+\rootk)}\,,
\end{align}
which is non-vanishing for all values of $\xi_2\leq 1/(16\zh)$ and therefore signals a curvature singularity at the event horizon. To avoid the latter, we impose $\theta_1=0$, which also leads to a well defined derivative $h^{(1)}(\zh)$:
\begin{align}
h^{(1)}(\zh)=\frac{a_2-\theta_2}{a_2\zh}\,,\label{h1hSerExp}
\end{align}
Moreover, the condition $\theta_1=0$ also changes the series expansion in (\ref{RicciSeriesExpansionThetaneq})
\begin{align}
&R=\frac{4\sqrt{2}\sqrt{\zh(1+\rootk)}\left((1+3\rootk)\theta_3+2\xi_3\right)}{(1+3\rootk)(1-8\zh\theta_2+\rootk)}\,(z-\zh)^{-1/2}+\mathcal{O}((z-\zh)^{0})\,.
\end{align}
We have furthermore verified that the singularity of $R$ at $z=\zh$ cannot be removed if $1-8\zh\theta_2+\rootk=0$ and we, therefore, require $\theta_2\neq a_2$.\footnote{This condition is compatible with $\theta_2<\frac{1+\varpi}{8\zh}$ which guarantees that $h^{(1)}(\zh)>0$ in (\ref{h1hSerExp}). The latter is necessary such that $h(z)>0$ for $z>\zh$ (with a simple zero at $z=\zh$).} In this case, the necessary condition for regularity of $R$ at the horizon is $(1+3\rootk)\theta_3+2\xi_3=0$. We have also verified that under the same condition also the Kretschmann scalar is finite at the horizon. To summarise, the consistency conditions for the approach outlined in Section~\ref{Sect:SeriesExpansion}, the conditions for the absence of a singularity of the Ricci scalar, and the bounds for positive metric functions $f$ and $h$ for $z>\zh$ are therefore
\begin{tcolorbox}[ams align,colback=black!10!white,colframe=black!95!green]
&\xi_1=0\,,&&\theta_1=0\,,&&\xi_3=-\frac{1}{2}\,(1+3\rootk)\,\theta_3\,,&&\xi_2\leq \frac{1}{16\zh}\,,&&\theta_2< \frac{1+\rootk}{8\zh}\,.
\label{ConRicciFinite}
\end{tcolorbox}
As remarked previously (see (\ref{ConditionsfhRegular})), the regularity condition (\ref{constraints}) found in the previous Section is compatible with this result and is the particular case $\xi_3=\theta_3=0$.

We further remark that the series coefficients (\ref{RecursionP}) and (\ref{Coeffbina}) along with the expansions (\ref{fhSerRho}) allow to compute the (finite) value of $R$ at the horizon. While the general form is rather complicated, here we only give the expression in the particular case $f=h$ (\emph{i.e.} $\xi_n=\theta_n$ $\forall n\geq 1$) with $\xi_1=\xi_3=0$ 
\begin{align}
R\big|_{z=\zh}=\frac{(1-\rootk^2)(a_2^2-a_4 \zh)}{8 \zh^3 a_2^3}+\frac{2\xi_4}{\zh a_2^2}=\frac{3+(2-5\rootk)\rootk}{2\zh^2(1+2\rootk)}+\frac{192\zh \xi_4}{1+3\rootk+2\rootk^2}\,.
\end{align}
which we shall use in the examples of the following Section.

Before closing this Section, we also provide the expression for the Hawking temperature
\begin{align}
T_{\text{H}}=\frac{\sqrt{\fn{1}h_H^{(1)}}}{4\pi}=\frac{\sqrt{1+\varpi-8\zh \theta_2}}{4\pi\sqrt{2}\,\zh}\,,
\end{align}
which agrees with the expression (\ref{Tempkappa}) found in the previous Section. The upper bound on $\theta_2$ in (\ref{ConRicciFinite}) guarantees that $T_{\text{H}}>0$. Indeed, $T_{\text{H}}=0$ would require $\fn{1}=0$ and/or $h_H^{(1)}=0$, which translate into $a_2=\xi_2$ and $a_2=\theta_2$ respectively. The former has no real solution, while the latter leads to a singularity of the Ricci scalar at the horizon. We remark, however, that black hole solutions with $T_{\text{H}}=0$ are possible upon choosing the solution $a_2=\frac{1-\varpi}{8\zh}$ in (\ref{Coefs2Extract}).

\section{Examples}\label{examples}
To illustrate further the approach presented in the previous Sections, in particular, the conditions (\ref{constraints}) and (\ref{ConRicciFinite}) which are sufficient for a finite Ricci scalar at the horizon, we shall consider two concrete examples: the first one is the Bonanno-Reuter \cite{Bonanno:2000ep} black hole, while the second one is specifically constructed to satisfy (\ref{constraints}) in a minimal fashion. Further examples from the literature are discussed in Appendix~\ref{Sect:FurtherExamples}.


\subsection{Example 1: Bonanno-Reuter asymptotically safe black hole}\label{Sect:ExampleBonannoReuter}
We first consider as an example the black hole metric introduced by Bonanno-Reuter \cite{Bonanno:2000ep} as a renormalization group improved generalization of the Schwarzschild space-time. Indeed, in this work, it has been proposed to replace the (dimensionful) Newton constant $G_{\text{Newton}}$ by a running Newton constant
\begin{align}
G_{\text{Newton}}\longrightarrow G(k)=\frac{G(k=0)}{1+\omega\,G(k=0)\,k^2} \,,
\end{align}
where $\omega\in\mathbb{R}$ is a constant and $k$ a (position dependent) scale (with reference scale $k=0$). The choice of the latter is ambiguous, but it has been proposed in \cite{Bonanno:2000ep} to use an inverse physical distance from the center of the black hole\footnote{Other options discussed in \cite{Bonanno:2000ep} include distances of the form $\int_{\mathcal{C}}\sqrt{|ds^2|}$, for different choices of contours $\mathcal{C}$ (for example the world-line of a free-falling observer). We shall discuss in future work (see also \cite{DAlise:2023hls}) that different such choices correspond to different schemes from the perspective of the renormalization group approach.}
\begin{align}
k(z)=\xi/\mathfrak{d}(z)\,,
\end{align}
with $\xi$ a suitable (dimensionful) constant. Such a distance has physical meaning, independent of a specific choice of coordinates. Adapting to our notation, the metric proposed in \cite{Bonanno:2000ep} can be written in the form (\ref{eq: metric}) with 
\begin{align}
    h(z)=f(z)=f_{\text{BR}}(z):=1-\frac{2\chi}{z}\frac{1}{1+\widetilde{\omega}/\mathfrak{d}(z)^2} \, ,\label{BRmetricfunction}
\end{align}
where $\widetilde{\omega}=\omega\xi^2$ is a dimensionless constant. In \cite{Bonanno:2000ep} the concrete value $\widetilde{\omega}=\frac{118}{15\pi}$ was given, however, subsequent works in the literature \cite{Donoghue:1993eb,Donoghue:1994dn,Hamber:1995cq,Bjerrum-Bohr:2002gqz,Donoghue:2022chi} potentially point towards different values (and a different sign). In the following, we shall consider $\widetilde{\omega}$ a generic parameter and discover marked differences between positive and negative values. Furthermore, we shall consider (\ref{BRmetricfunction}) to be valid only outside of the event horizon, which is located at $d_{\text{BH},H}$ (which we take as an input of the model.\footnote{Here we are allowing for the possibility that the metric inside of the black hole is different from (\ref{BRmetricfunction}) in which case $d_{\text{BH},H}$ would need to be computed as a separate input to the model. As we shall see, our conclusions will be entirely independent of this choice and thus the concrete value of $d_{\text{BH},H}$.}) In order for $f$ to remain well defined at the horizon, we shall assume $\widetilde{\omega}+d^2_{\text{BR},H}=\frac{2\chi d_{\text{BR},H}^2}{\zh}\neq 0$.

For concrete computations, a choice for the physical distance needs to be made. Here we shall discuss three different possibilities that lead to a geometry with an (outer) event horizon, for which we can verify whether the conditions (\ref{constraints}) are satisfied and whether therefore the Ricci scalar is finite, namely: \emph{(i)} the proper distance computed from the metric~(\ref{eq: metric}); \emph{(ii)} the proper distance computed from the Schwarzschild metric; \emph{(iii)} an interpolating function. We shall discuss all three possibilities in the following:
\begin{itemize}
\item[\emph{(i)}] choosing $\mathfrak{d}(z)$ as the \emph{proper distance} $d(z)$ in eq.~(\ref{propd}):
\begin{align}
&d_{\text{BR}}(z)=\int_0^z \frac{1}{\sqrt{|f_{\text{BR}}(z)|}}\,,&&\forall z\geq 0\,.\label{BRselfDistance}
\end{align}
This is a self-consistent choice in the sense that the proper distance is compatible with the metric~(\ref{eq: metric}). As discussed in \cite{Binetti:2022xdi}, this guarantees that the modified metric exhibits the same diffeomorphism invariance as the (classical) Schwarzschild black hole. However, explicitly computing the distance becomes more involved (since (\ref{BRselfDistance}) is an implicit definition). For negative values of $\widetilde{\omega}$, a series expansion close to the horizon is developed in Appendix~\ref{Sect:BRDistanceComputations}. However, for our purposes, this is not in fact required, since we can simply verify the regularity conditions developed in the previous Sections, \emph{i.e.} eq.~(\ref{constraints}) or (\ref{ConRicciFinite}). For the concrete function~(\ref{BRmetricfunction}) with $\mathfrak{d}(z)=d_{\text{BR}}(z)=d_{\text{BR},H}+\rho(z)$, we obtain
\begin{align}
&\fid{1} =\pid{1}=
     - \frac{2 \widetilde{\omega}}{d_{\text{BR},H}(d_{\text{BR},H}^2 + \widetilde{\omega})} \neq 0\ ,&&\xi_1=\theta_1=\frac{4\chi \widetilde{\omega} d_{\text{BR},H}}{(\widetilde{\omega}+d_{\text{BR},H}^2)^2}\neq 0\,,\label{BRbreakingRelation}
\end{align}
such that neither (\ref{ConRicciFinite}) nor (\ref{constraints}) are satisfied. In fact, (\ref{BRbreakingRelation}) implies that already the first derivative of $f_{\text{BR}}$
\begin{align}
    f_{\text{BR}}^{(1)}(z)=(1-f_{\text{BR}}(z))\left(\frac{1}{z}-\frac{2 \widetilde{\omega}}{d_{\text{BR}}(z)(d_{\text{BR}}(z)^2+\widetilde{\omega})}\frac{1}{\sqrt{f_{\text{BR}}(z)}} \right)\,,&&\forall z\geq z_{\text{BR},H}\,,\nonumber
\end{align}
diverges at the horizon
\begin{align}
\lim_{\epsilon\to 0^+}  f_{\text{BR}}^{(1)}(z_{\text{BR},H}+\epsilon)\to \infty\,.
\end{align}
This is due to the fact that $f_{\text{BR}}(z_{\text{BR},H})=0$. Following the discussion of Section~\ref{Sect:ThermoGeneral}, this poses problems with the interpretation of the black hole's thermodynamical properties, notably the Hawking temperature's definition. Furthermore, it also leads to a curvature singularity at the horizon, since for example, the Ricci scalar becomes 
\begin{align}
     &R= \frac{\widetilde{\omega}  (f_{\text{BR}}-1) \left( d_{\text{BR}}(d_{\text{BR}}^2+ \widetilde{\omega}) (1-5 f_{\text{BR}})f_{\text{BR}}^{1/2}  +6 z d_{\text{BR}}^2 f_{\text{BR}}-2 \widetilde{\omega}  z\right)}{z d_{\text{BR}}^2  \left(d_{\text{BR}}^2+ \widetilde{\omega}\right)^2 f_{\text{BR}}^2} \ ,&&\forall z\geq \zh\,.\nonumber
\end{align}
This expression diverges at the event horizon due to the factor $f_{\text{BR}}^{2}$ in the denominator (while the numerator at the horizon assumes the finite value $2\widetilde{\omega}^2 z_H$).

Finally, we remark since $\xi_1\neq 0$, the results of the series expansion approach developed in Section~\ref{Sect:SeriesExpansionApproach} are not directly applicable. In appendix~\ref{App:SerNonZeroXi}, we show how it can be generalised in this case, and the consequences for the Bonanno-Reuter space-time for $\widetilde{\omega}<0$, are discussed in appendix~\ref{Sect:BRDistanceComputations} (confirming further our above conclusions in this case).

\item[\emph{(ii)}] choosing $\mathfrak{d}(z)$ as the \emph{proper distance} of the (classical) \emph{Schwarzschild geometry}, \emph{i.e.}
\begin{align}
d_{\text{S}}(z)=\int_0^z\frac{dz'}{\sqrt{\left|1-\frac{2\chi}{z}\right|}}=\left\{\begin{array}{lcl}\pi\chi-\sqrt{z(2\chi-z)}-2\chi \arctan\sqrt{\frac{2\chi}{z}-1} & \text{if} & 0<z<2\chi\,, \\[2pt] \pi\chi+\sqrt{z(z-2\chi)}+2\chi \text{arctanh}\sqrt{1-\frac{2\chi}{z}} & \text{if} & 2\chi<z\,. \end{array}\right.\label{SchwarzschildDistance}
\end{align}
This option was initially advocated in \cite{Bonanno:2000ep} and has the advantage that it can be computed as a closed expression in terms of $z$. This geometry possesses a horizon, whose position is corrected by $\widetilde{\omega}$, \emph{e.g.} for large mass $\chi$ of the black hole
\begin{align}
z_{\text{BR},H}=2\chi-\frac{2}{\pi^2}\,\frac{\widetilde{\omega}}{\chi}+\mathcal{O}\left(|\widetilde{\omega}|^{3/2}/\chi^2\right)\,.
\end{align}
For $f_{\text{BR}}$ to be well defined at the horizon, we assume that $\widetilde{\omega}+d_{\text{S}}^2(z=z_{\text{BR},H})\neq 0$. The derivative of $f$ in this case becomes
\begin{align}
f^{(1)}_{\text{BR}}(z)=\frac{2\chi}{z^2\left(1+\frac{\omega}{d_{\text{S}}^2}\right)}-\frac{4\chi \omega}{z\,d_{\text{S}}^3 \left|1-\frac{2\chi}{z}\right| \left(1+\frac{\omega}{d_{\text{S}}^2}\right)^2}\,,
\end{align}
which is finite at the horizon $z_{\text{BR},H}$, but diverges for $z=2\chi$ (which for $\widetilde{\omega}>0$ lies outside the horizon of the black hole). Because of this, while the Ricci scalar is finite at $z=z_{\text{BR},H}$, the geometry has a curvature singularity at $z=2\chi$.

\item[\emph{(iii)}] Although a closed function of $z$, the Schwarzschild proper distance (\ref{SchwarzschildDistance}) is still difficult to work with for concrete computations. Therefore in \cite{Bonanno:2000ep} the following approximation for (\ref{SchwarzschildDistance}) was proposed\footnote{We are using a notation adapted to the current paper.}
\begin{equation}
d_{\text{S}}(z)\simeq \kappa_{\text{BR}}(z)=\left(\frac{z^3}{z+\gamma\,\chi}\right)^{1/2}\,,\qq{such that} \begin{array}{l} \displaystyle \lim_{z\to 0} \kappa_{\text{BR}}(z)\sim \frac{z^{3/2}}{\sqrt{\gamma \chi}}\,,\\ \displaystyle\lim_{z\to\infty} \kappa_{\text{BR}}(z)\sim z\,. \end{array}\label{BRdistanceApprox}
\end{equation}  
Here $\gamma\in\mathbb{R}$ is a constant, which in order to mimic the same behaviour as (\ref{SchwarzschildDistance}) at the origin ($z\to 0$) needs to be chosen as $\gamma=9/2$. The function $\kappa_{\text{BR}}$ has no inflection points, and thus does not feature the same behaviour as a proper distance at a horizon for any value of $z$. 

Identifying $\mathfrak{d}$ in (\ref{BRmetricfunction}) with $\kappa_{\text{BR}}$ leads to a zero of $f_{\text{BR}}$ at 
\begin{align}
z_{\text{BR},H}=2\chi-\frac{2+\gamma}{4}\,\frac{\omega}{\chi}+\mathcal{O}(\omega^2/\chi^3)\,.
\end{align} 
At this horizon, both the first and second derivatives of $f_{\text{BR}}$ are finite implying that the Ricci scalar takes a finite value.

\end{itemize}
To summarise, treating (\ref{BRmetricfunction}) in a \emph{self-consistent fashion} by identifying $\mathfrak{d}$ by the proper distance calculated from $f_{\text{BR}}$ itself leads to a divergent first derivative of the metric function at the horizon, which in turn leads to significant problems for physical quantities. Notably, it poses problems for defining a finite Hawking temperature and leads to a curvature singularity at the (outer) event horizon. This is in line with the results of Sections~\ref{Sect:Conditions} and \ref{Sect:SeriesExpansionApproach}, due to the fact that the function $\frac{2\chi}{1+\widetilde{\omega}/(d_{\text{BR},H}+\rho)^2}$ does not satisfy the conditions (\ref{constraints}), independent of the geometry of the black hole inside the event horizon (\emph{i.e.} independent of the value of $d_{\text{BR},H}$). The choices \emph{(ii)} and \emph{(iii)} constitute a departure from the original idea presented in \cite{Bonanno:2000ep} (proposed as an approximation in this work) by replacing $\mathfrak{d}$ by a function of the radial coordinate $z$, which does \emph{not} represent a physical distance that has been consistently calculated from the metric characterised by (\ref{BRmetricfunction}). 

While \emph{(ii)}, depending on the sign of $\widetilde{\omega}$, may have a curvature singularity outside of the event horizon, choice \emph{(iii)} is at least well-behaved from this perspective. However, from the point of view of the original motivation to deform the Schwarzschild metric by a function of a (consistently calculated) distance function, the choice \emph{(iii)} corresponds to a different deformation function than (\ref{BRmetricfunction}). Using the approach outlined in Section~\ref{Sect:SeriesExpansionApproach}, we can determine this modified deformation function by reverse engineering the coefficients $\xi_n$: indeed, by integrating eq.~(\ref{SeriesForm}) we find the coefficients $b_n$ in eq.~(\ref{SeriesRho})
\begin{align}
&b_1=\frac{\sqrt{2}|\zbr^3+\widetilde{\omega}(\zbr+\gamma\chi)|}{\sqrt{\zbr\chi(\zbr^3-\widetilde{\omega}(\zbr+2\gamma\chi))}}\,,\hspace{1cm}b_2=0\,,\nonumber\\
&b_3=\frac{b_1^3\chi\left(\zbr^6-3\zbr^4\widetilde{\omega}-7\zbr^3\gamma\chi\widetilde{\omega}+\gamma^2\chi^2\widetilde{\omega}^2\right)}{24\sqrt{2}\left(\zbr^3+\widetilde{\omega}(\gamma\chi+\zbr)\right)^3}\,,\hspace{1cm}b_4=0\,.
\end{align}
Therefore with
\begin{align}
&a_1=0\,,&&a_2=\frac{1}{b_1^2}\,,&&a_3=0\,,&&a_4=-\frac{2b_3}{b_1^5}\,,
\end{align}
we obtain for the leading coefficients of $\xi_n$ (which are equal to $\theta_n$) 
\begin{align}
&\xi_1=0\,,&&\xi_2=\frac{2\zbr^2\chi(2\zbr+3\gamma\chi)\widetilde{\omega}}{b_1^2(\zbr^3+\widetilde{\omega}(\zbr+\gamma\chi))^2}\,,&&\xi_3=0\,.
\end{align}
These indeed satisfy the conditions (\ref{ConRicciFinite}). Therefore, the ``approximation'' to use (\ref{BRdistanceApprox}) for $\mathfrak{d}$ in (\ref{BRmetricfunction}) instead of the self-consistently calculated proper distance, corresponds to changing the metric function (\ref{BRmetricfunction}), in a way characterised by the above expansion coefficients.

\subsection{Example 2: Minimal metric deformation}
\begin{center}
    \begin{figure}[!t]
        \centering
        \includegraphics[width=0.75\textwidth]{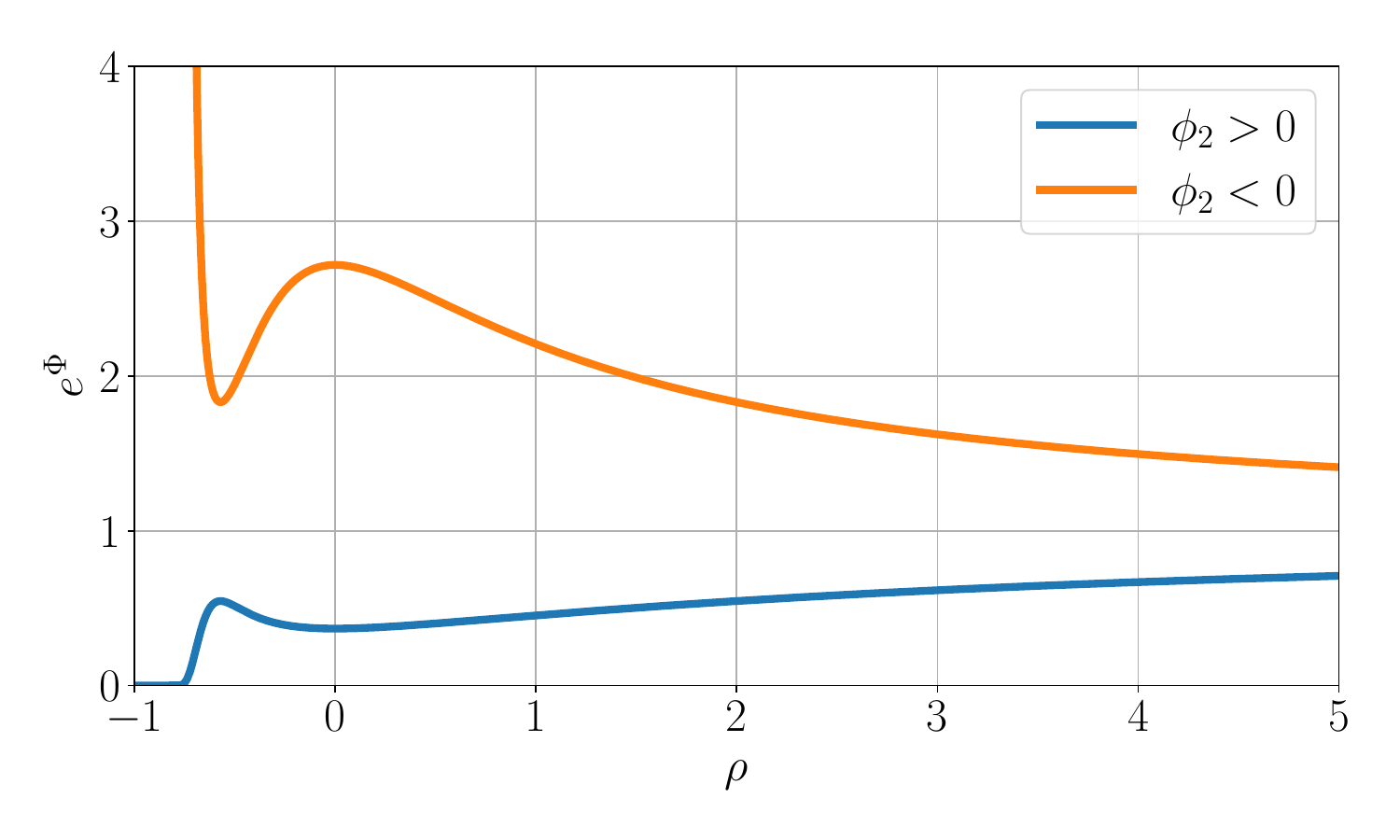}
        \caption{\emph{Function $e^{\Phi}$ in eq.~\eqref{FormExpPhiExample} with $\phi_0=-3\phi_2/2\dho^2$ and $\dho=1$.}}
        \label{Fig:MinimalExample}
    \end{figure}
\end{center}

As a further (novel) example, we consider the following minimal solution of the conditions~\eqref{constraints}
\begin{align}
&\Phi_H=\Psi_H=\phi_0\,,&&\fid{1}=\pid{1}=0\,,&&\fid{2}=\pid{2}=\phi_2\,,&&\fid{3}=\pid{3}=-6\dho \phi_2\, ,\label{Phifuns}
\end{align}
and $\fid{n}=0$ $\forall n\geq 4$. Here $\phi_0,\phi_2\in\mathbb{R}$ are arbitrary parameters, which, however, are not independent: indeed, in order for the metric to asymptotically, approach the Schwarzschild one (with mass parameter $\chi$), we require the asymptotic limit 
\begin{align}
\lim_{\rho\to\infty}e^{\Phi\left(\frac{1}{\dho+\rho}\right)}=e^{\phi_0+\frac{3\phi_2}{2\dho^2}}\overset{!}{=}  1 \quad \implies  \quad \phi_0 = -\frac{3\phi_2}{2\dho^2}\ .\label{ExampleLimits}
\end{align}
Here we also consider $\dho$ as a parameter of the model, which encodes information about the interior of the black hole. With this, the solution (\ref{Phifuns}) of~\eqref{constraints} can be written compactly in the form
\begin{align}
\Phi\left(\frac{1}{\dho+\rho}\right)=\Psi\left(\frac{1}{\dho+\rho}\right)=-\frac{3\phi_2}{2\dho^2}+\frac{\rho^2(\dho+3\rho)\phi_2}{2\dho^2(\dho+\rho)^3}\,.\label{FormExpPhiExample}
\end{align}
Choosing $\fid{n}\neq 0$ (or $\pid{n}\neq 0$) for $n\geq 4$ in (\ref{Phifuns}) would yield higher modifications of $e^{\Phi}$ of order $\mathcal{O}(\rho^4)$, which are negligible close to the horizon. The position of the latter is located at 
\begin{align}
\zh=2\chi\, e^{\phi_0}=2\chi\,\text{exp}\left(-\frac{3\phi_2}{2\dho^2}\right)\,.
\end{align}
A schematic plot of $e^\Phi$ as a function of $\rho$ is shown in Figure~\ref{Fig:MinimalExample}. Here we have plotted $e^\Phi$ in the entire space-time (\emph{i.e.} also inside of the event horizon for $\rho<0$): following our general philosophy, we shall discuss the metric function (and all associated quantities) only outside of the horizon (\emph{i.e.} for $\rho\geq 0$) and only make a few brief remarks on the physics inside of the black hole in Section~\ref{Sect:MinimalModelInside}.
The coefficients $\kappa_n$ in a series expansion of $\Phi$ in powers of $\rho$ (see eq.~(\ref{SerPhiKappa})) are worked out in eq.~(\ref{MinExCoefsKappa}), while the leading coefficients $\xi_n$ (stemming from the series expansion of $2\chi\,e^\Phi$ in eq.~(\ref{XiCoefs})) are exhibited in (\ref{CoefsMinModel}). In particular, due to the choice (\ref{Phifuns}), we find $\xi_1=0=\xi_3$, which therefore satisfies (\ref{constraints}) and (\ref{ConRicciFinite}). Moreover, since (\ref{Phifuns}) also implies $\Phi=\Psi$ (and thus $\xi_n=\theta_n$), the condition (\ref{ConRicciFinite}) is also trivially satisfied, such that we expect the Ricci scalar of this model to be finite at the event horizon $\zh$.


\subsubsection{Curvature and temperature}
As a first step to calculating physical quantities for the space-time metric characterised by (\ref{Phifuns}), we compute the derivative (\ref{f1h h1h}) of the functions $f$ and $h$ at the horizon $\zh=2\chi e^{\phi_0}$
\begin{align}
&\fn{1}=\hn{1}=\frac{e^{-\phi_0}}{4\chi}\left(1+\sqrt{1-\frac{32 e^{2\phi_0}\chi^2\phi_2}{\dho^4}}\right) \qq{with} \phi_0=-\frac{3\phi_2}{2\dho^2}\,,\label{ModelDerivativefh}
\end{align}
which imposes the condition $\phi_2\leq \frac{\dho^4}{32\,e^{2\phi_0}\,\chi^2}$, which with (\ref{ExampleLimits}) therefore becomes the non-linear relation for $\phi_2$
\begin{align}
\phi_2\,e^{3\phi_2/\dho^2}\leq \frac{\dho^4}{32\chi^2}\,.
\end{align}
The Ricci scalar at the horizon is finite in this model and takes the value
\begin{align}
&R\big|_{\zh}=\frac{3+(2-5\rootk)\rootk}{2\zh^2(1+2\rootk)}+\frac{48 e^{-\frac{3\phi_2}{2\dho^2}}\zh\chi \phi_2(\phi_2-12\dho^2)}{\dho^8(1+\rootk)(1+2\rootk)}\,,&&\text{with} &&\rootk=\sqrt{1-\frac{16 \zh \phi_2 \chi e^{-\frac{3\phi_2}{2\dho^2}}}{\dho^4}}\,.
\end{align}

Finiteness of the derivatives (\ref{ModelDerivativefh}) (at the event horizon) is a necessary requirement for well-behaved thermodynamical properties of the black hole. Indeed, the Hawking temperature is given by
\begin{equation}\label{ExampleTemp}
    T_\mathrm{H}=\frac{\fn{1}}{4\pi}=\frac{1}{8 \pi \zh}\left(1+\sqrt{1-\frac{8 \zh^2\phi_2}{\dho^4}}\right)\ .
\end{equation}
Determining the entropy using eq.~(\ref{EntropyDef}) requires specifying the $\chi$-dependence of $\dho$ and $\phi_2$ and thus requires further refinement of the model. To give a concrete example, we shall consider $\chi\gg 1$ along with  
\begin{align}
&\phi_2=\mathcal{O}(\chi^0) \qq{and} \dho=\pi\chi+ \mathcal{O}(\chi^0)\,,\label{InteriorMassDependence}
\end{align}
that is, we are assuming only subleading corrections to the classical Schwarzschild metric. We, therefore, find the horizon position
\begin{align}
\zh=2\chi-\frac{3\phi_2}{\pi^2 \chi}+\mathcal{O}(\chi^{-2})\,,
\end{align}
and thus for the Hawking temperature 
\begin{align}
T_\mathrm{H}=\frac{1}{8\pi \chi}\left[1+\frac{(3\pi^2 -16)\phi_2}{2\pi^4 \chi^2}+\mathcal{O}(\chi^{-3})\right]\,,
\end{align}
and the entropy
\begin{align}
    S = 4 \pi  \chi ^2 \left[1-\frac{ \left(3 \pi ^2-16\right) \phi _2 }{3 \pi^2 \chi^2 }\log (\chi^2 ) + \mathcal{O} \left( \frac{1}{\chi^4}\right)\right] + \mathrm{const}\ .\label{FormEntropyExample}
\end{align}
This approximation exhibits a logarithmic correction, compatible with previous results in the literature \cite{Fursaev:1994te, Xiao:2021zly, Solodukhin:2011gn, Solodukhin:1994yz, Sen:2012dw, El-Menoufi:2015cqw, Cai:2009ua, Carlip:2000nv, Banerjee:2008cf, Banerjee:2008fz, Kaul:2000kf}. We remind the reader, however, that (\ref{FormEntropyExample}) is based on the assumptions (\ref{InteriorMassDependence}), which are related to the interior of the black hole solution. 

\subsubsection{Extending the metric inside the horizon}\label{Sect:MinimalModelInside}
Following our general philosophy, so far we have considered the metric function only outside of the black hole horizon and have used the horizon distance $\dho$ (and $\zh$) as the only additional input that is required for the interior of the black hole. Indeed, due to its definition (\ref{propd}), the calculation of $\dho$ requires knowing the metric in the interior of the black hole, which we have not specified up to this point. In the current example, since the function (\ref{FormExpPhiExample}) can in fact be extended to the entire space-time (\emph{i.e.} also for $\rho<0$), as is showcased in Figure~\ref{Fig:MinimalExample}, one can contemplate the possibility to use (\ref{FormExpPhiExample}) as a model for the entire space-time. Although this discussion is generally outside of the scope of this paper, here we shall nevertheless make a few remarks regarding this possibility. For concreteness, we shall focus on $\phi_2>0$ in the entire Subsubsection.

As a first question, we may ask the behavior of the metric at the origin. In order to get a better understanding of the space-time described by (\ref{FormExpPhiExample}) for $z\ll 1$, we may use a right rectangular approximation for the integral of the proper distance (\ref{propd})
\begin{align}
&d(z)\simeq \frac{z}{\sqrt{|f(z)|}} \qquad \forall z\ll 1\,,\label{ApproximationOrigin}
\end{align}
which is an algebraic equation. Numerical analysis suggests that there are no real positive values for $(z,d(z))$ that satisfy this relation for small $z$ when $f(z)<0$. We have, however, obtained the following solution of (\ref{ApproximationOrigin}) for $z(d)$ in the case $f(z)>0$
\begin{align}
\frac{z}{d}&=\frac{\sqrt{d}\left(3^{1/3}d^{2/3}+\left(-9e^{\Phi}\chi+\sqrt{-3d^2+81 e^{2\Phi}\chi^2}\right)\right)^{2/3}}{3^{2/3}\left(-9e^{\Phi}\chi+\sqrt{-3d^2+81e^{2\Phi}\chi^2}\right)^{1/3}}\,\nonumber\\
&=\cos\left[\frac{1}{3}\,\arcsin\left(\frac{3\sqrt{3} e^{\Phi}\chi}{d}\right)\right]-\frac{1}{\sqrt{3}}\sin\left[\frac{1}{3}\,\arcsin\left(\frac{3\sqrt{3} e^{\Phi}\chi}{d}\right)\right]\,.\label{zSerDef1}
\end{align}
The function $f$, therefore, approaches $+1$ for $d\to 0$, which hints towards the absence of a singularity at the origin. To verify this, we realize that the function $z$ in (\ref{zSerDef1}) cannot be expanded in a Taylor series for $d$. However, noting that (for $\Phi$ given in (\ref{FormExpPhiExample})) for $\phi_2>0$ we have the limit $\displaystyle \lim_{\rho\to -\dho}\,e^{\Phi}$ in an exponential fashion, we can (formally) write $z/d$ as a series expansion in powers of $e^\Phi$
\begin{align}
&z/d=1-\sum_{n=1}^\infty \frac{2^{n-1}\,\Gamma\left(\frac{3n-1}{2}\right)}{n!\,\Gamma\left(\frac{n+1}{2}\right)}\,\left(\frac{\chi e^\Phi}{d}\right)^n \qq{for} \frac{\chi\,e^\Phi}{d}\in\left[0,\tfrac{1}{3\sqrt{3}}\right)\,.\label{Taylorzd}
\end{align}
We then find the Ricci scalar close to the origin 
\begin{align}
R=\frac{2\chi \,e^\Phi\, \phi_2}{\dho^2 d^9}\left[\dho^2 d^3(7d-6\dho)+(3\dho-4d)^2(d-\dho)^2\phi_2\right]+\mathcal{O}(e^{2\Phi})\,,\label{RicciOrigin}
\end{align}
which is indeed finite in the limit $d\to 0$, due to the exponential suppression of $e^\Phi$.\footnote{We have also verified that the Kretschmann scalar is finite at the origin in this model.}

While the above result of the absence of a curvature singularity at the origin is very encouraging conceptually, it also highlights another problem: indeed, if $f(z)>0$ close to the origin and $f(z)>0$ for $z>\zh$, with a simple zero at $\zh$, $f$ necessarily has (at least) one further zero in the interval $z\in(0,\zh)$, more concretely for $\rho\leq -\dho/3$. In other words, the space-time described by (\ref{FormExpPhiExample}) has at least one more inner horizon. At the latter, we have to verify again if all necessary conditions are met for the absence of a curvature singularity. However, since (\ref{Phifuns}) is only a minimal solution of the conditions (\ref{constraints}) (which are tailored to remove unphysical singularities at $\rho=0$), we cannot guarantee the absence of a curvature singularity at this inner horizon. This problem can be circumvented by allowing some of the $\fid{n}$ (for $n\geq 4$) to be non-zero. While we leave a more detailed discussion for the black hole interior to further work, we show in Appendix~\ref{App:InnerHorizon} how to derive conditions for the absence of curvature singularities at an inner horizon.

\section{Large distance expansion} 
\label{LDE}
\begin{figure}[h!]
    \begin{center}
        
\tikzset{every picture/.style={line width=0.75pt}} 

\begin{tikzpicture}[x=0.75pt,y=0.75pt,yscale=-1,xscale=1]

\draw  [color={rgb, 255:red, 0; green, 0; blue, 255 }  ,draw opacity=1 ][fill={rgb, 255:red, 105; green, 105; blue, 252 }  ,fill opacity=1 ] (93.99,64.29) -- (353.82,64.29) -- (353.82,96.33) -- (93.99,96.33) -- cycle ;
\draw  [color={rgb, 255:red, 6; green, 166; blue, 6 }  ,draw opacity=1 ][fill={rgb, 255:red, 121; green, 213; blue, 121 }  ,fill opacity=1 ] (96.37,69.92) -- (180.6,69.92) -- (180.6,91.57) -- (96.37,91.57) -- cycle ;
\draw [line width=1.5]  [dash pattern={on 5.63pt off 4.5pt}]  (83.16,37.44) .. controls (92.04,54.54) and (93.77,60.61) .. (93.99,80.74) ;
\draw [line width=1.5]  [dash pattern={on 5.63pt off 4.5pt}]  (83.16,124.05) .. controls (92.04,106.94) and (93.77,100.88) .. (93.99,80.74) ;

\draw  [color={rgb, 255:red, 208; green, 2; blue, 27 }  ,draw opacity=1 ][fill={rgb, 255:red, 229; green, 101; blue, 112 }  ,fill opacity=1 ] (244.58,69.92) -- (350.03,69.92) -- (350.03,91.57) -- (244.58,91.57) -- cycle ;
\draw [line width=1.5]    (7.37,80.74) -- (340,80.74) ;
\draw [shift={(343,80.74)}, rotate = 180] [color={rgb, 255:red, 0; green, 0; blue, 0 }  ][line width=1.5]    (9.95,-4.46) .. controls (6.32,-2.09) and (3.01,-0.61) .. (0,0) .. controls (3.01,0.61) and (6.32,2.09) .. (9.95,4.46)   ;
\draw [shift={(7.37,80.74)}, rotate = 0] [color={rgb, 255:red, 0; green, 0; blue, 0 }  ][fill={rgb, 255:red, 0; green, 0; blue, 0 }  ][line width=1.5]      (0, 0) circle [x radius= 3.05, y radius= 3.05]   ;

\draw (365,72) node [anchor=north west][inner sep=0.75pt]  [font=\normalsize]  {$z,\ d,\ \rho $};
\draw (235.31,104) node [anchor=north west][inner sep=0.75pt]  [font=\normalsize,color={rgb, 255:red, 208; green, 2; blue, 27 }  ,opacity=1 ]  {$\displaystyle 2\chi \left( 1+\sum _{n=1}^{\infty }\frac{\omega _{n}}{d^{n}}\right)$};
\draw (110,104) node [anchor=north west][inner sep=0.75pt]  [font=\normalsize,color={rgb, 255:red, 6; green, 166; blue, 6 }  ,opacity=1 ]  {$\displaystyle  \sum _{n=0}^{\infty } \xi_{n} \rho ^{n}$};
\draw (180.4,35) node [anchor=north west][inner sep=0.75pt]  [font=\normalsize,color={rgb, 255:red, 0; green, 0; blue, 255 }  ,opacity=1 ]  {$\displaystyle 2 \chi e^{\Phi ( 1/d)}$};
\draw (65,-10) node [anchor=north west][inner sep=0.75pt]  [font=\normalsize]  {$ \begin{array}{c}
\zh ,\dho \\
\rho = 0
\end{array}$};

\end{tikzpicture}

    \end{center}
    \caption{\emph{Schematic division of the space outside of the black hole in three main regions. The curved dashed line represents the position of the event horizon. The blue region is where the \emph{full non-perturbative} form of the metric functions, given in eq.~\eqref{modifiedfh} is expected to hold (\emph{i.e.} the whole space-time outside the event horizon). For simplicity, we only give the expressions for $f$ in the Figure, but a similar form also holds for $h$. The green region is where the metric is expanded as a convergent series in the proper distance from the event horizon $\rho$, as given in eqs.~\eqref{XiCoefs} and~\eqref{ThetaCoefs}. Finally, the red region refers to the asymptotically large distances from the event horizon, where the metric can be expanded in inverse powers of the proper distance, as given in eqs~\eqref{Functionsfh}.}}
    \label{ranges}
\end{figure}
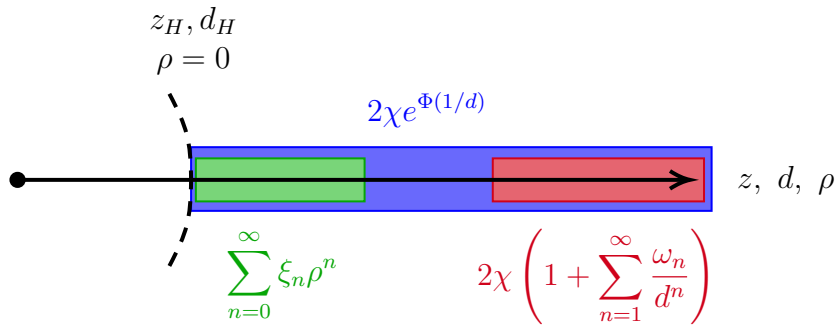

So far, we have focussed on the consistency
 conditions arising at and near the event horizon for generic static and spherically symmetric black hole metrics. More generally we have investigated the impact of these conditions on an effective metric expanded in terms of a physical distance from the event horizon. The range of applicability of this theory is visualized via the green box in Fig.~\ref{ranges}. The generic metric \eqref{eq: metric} in \eqref{modifiedfh}  covers any distance from the black hole horizon as represented in the blue box.  
 
We now investigate an asymptotic expansion from an infinite distance valid in the red box. This was considered in \cite{Binetti:2022xdi} and extended in \cite{DAlise:2023hls}:
\begin{equation}
f(z)=1-\frac{2\chi}{z}\left(1+\sum_{n=1}^\infty\ \frac{\omega_n}{d(z)^{n}}\right) \qq{and}  h(z)=1-\frac{2\chi}{z}\left(1+\sum_{n=1}^\infty\ \frac{\gamma_n}{d(z)^{n}}\right) \ , \label{Functionsfh}
\end{equation}
where $\omega_n,\gamma_n$ are effective coefficients encoding the deformation from the Schwarzschild solution, and the physical distance $d(z)$ is defined in eq.~\eqref{propd}. For $n$ even and $\gamma_n=\omega_n$ we recover the metric and results of \cite{Binetti:2022xdi}. By construction, the expansion ensures that the classical black hole metric is recovered at an infinite distance while its radius of convergence depends on the $\omega_n$ and $\gamma_n$ coefficients.

\subsection{Convergence criteria and derivatives}
We now investigate the impact of the event horizon constraints (\ref{constraints}) and (\ref{ConRicciFinite}) on the large distance expansion coefficients $\omega_n$ and $\gamma_n$ under the assumption that (\ref{Functionsfh}) is convergent up to the horizon. In other words, we assume that the red box in Fig.~\ref{ranges} extends all the way to the event horizon. This imposes certain conditions on the coefficients $\omega_{n}$ and $\gamma_n$, namely that the radius of convergence of the series \eqref{Functionsfh} is larger than the inverse distance of the horizon
\begin{equation}
\limsup_{n\to\infty}|\omega_n|^{\frac{1}{n}}\leq \dho \qq{and} \limsup_{n\to\infty}|\gamma_n|^{\frac{1}{n}}\leq \dho\ ,\label{RadiusConv}
\end{equation}
where $\limsup$ denotes the limit superior. In the following, we shall find it useful to re-scale the coefficients $\omega_n$ and $\gamma_n$ by $\dho^n$
\footnote{For the purpose of a large $\chi$ expansion (see Section~\ref{Sect:Thermo}), one could equally rescale by powers of the classical horizon distance $\pi\chi$.}
\begin{equation}
\omega_n=\bar{\omega}_n\,\dho^n \qq{and} \gamma_n=\bar{\gamma}_n\,\dho^n\ , \quad \forall n\in\mathbb{N}\ ,\label{CoeffsRescale}
\end{equation}
where $\bar{\omega}_n$ and $\bar{\gamma}_n$ are coefficients of a series with radius of convergence $\geq 1$. 

The position of the horizon $z_H$ is determined by
\begin{equation}
   h(z_H) = 1 - \frac{2\chi}{\zh}\left(1+\sum_{n=1}^\infty\bar{\gamma}_n\right) =0 \qq{and} f(\zh) = 1 - \frac{2\chi}{\zh}\left(1+\sum_{n=1}^\infty \bar{\omega}_n\right)=0\ ,
\end{equation}
which leads to the following relation for the (convergent) series 
\begin{equation}\label{omega zh}
\sum_{n=1}^\infty\bar{\omega}_n=\sum_{n=1}^\infty\bar{\gamma}_n\ = \frac{\zh}{2\chi}-1\ .
\end{equation}
We also have
\begin{equation}
    \Phi_H= \log \sum_{n=0}^\infty \bar{\omega}_n\qq{and}
    \Psi_H = \log \sum_{n=0}^\infty \bar{\gamma}_n \ .
\end{equation}
The first derivative of $f$ for $z\geq \zh$ reads
\begin{align}
f^{(1)}(z):=\dv{f}{z} 
=\frac{1}{z}\left( 1-f(z) + \frac{2\chi}{\sqrt{{f(z)}}}\sum_{n=1}^\infty \frac{ n \omega_n}{d(z)^{n+1}} \right)\ . \label{fp}
\end{align}
Even assuming \eqref{RadiusConv}, such that the sum is convergent for all values of $d(z)$ up to the horizon,  this equation is still divergent at $z=\zh$ due to the $\sqrt{f(z)}$ in the denominator. A similar problem occurs for the first derivative of $h$. 
Concretely, expanding both derivatives $f^{(1)}$ and $h^{(1)}$ in terms of $\rho$ (similar to eq.~\eqref{fprimePhi}), we obtain
\begin{align}
    f^{(1)} 
&= \frac{1}{\zh}\left( 1 + \frac{4\chi}{\rho\,\dho \fn{1}}\sum_{n=1}^\infty  n\, \bar{\omega}_n  - \frac{2}{\dho^2 \fn{1}}\sum_{n=1}^\infty n (n+1)\, \bar{\omega}_n + \mathcal{O}(\rho) \right)\ ,\nonumber\\
h^{(1)}  &=  \frac{1}{\zh}\left( 1 + \frac{4\chi}{\rho\,\dho \fn{1}} \sum_{n=0}^\infty n\,\bar{\gamma}_n  - \frac{4\chi}{\dho^2 \fn{1} } \sum_{n=0}^\infty n(n+1)\,\bar{\gamma}_n + \mathcal{O}(\rho) \right)\ ,\label{fp rho}
\end{align}
which contain terms of order $\rho^{-1} \sim (z-\zh)^{-1/2} \sim f^{-1/2}$ that become divergent at the horizon. These can be removed by requiring
\begin{align}
&\sum_{n=1}^\infty   n\, \bar{\omega}_n = 0\,,&&\text{and} &&\sum_{n=0}^\infty n\,\bar{\gamma}_n
    =0
    \, . \label{omega cond 1}
\end{align}
From the remainder of \eqref{fp rho} we now find 
\begin{align}
&\fn{1} = \frac{1}{2 \zh} \left( 1 + \sqrt{1 - \frac{16 \zh \chi}{\dho^2} \sum_{n=1}^\infty n^2\, \bar{\omega}_n}\right)\,,&&\text{and} && h^{(1)}_H = \frac{1}{\zh} - \frac{4\chi}{\zh \dho^2 \fn{1}}\sum_{n=0}^\infty n^2\,\bar{\gamma}_n\ ,\label{fpH}
\end{align}
where we have again chosen a solution for $\fn{1}$ which corresponds to the Schwarzschild geometry for $\omega_{n}\to 0$ $\forall n\geq 1$. Comparing to the first equation in eq.~\eqref{f1h h1h}, we have
\begin{align}
&\fid{2} = \frac{2\chi \dho^2 }{\zh}\sum_{n=1}^\infty n^2\, \bar{\omega}_n\,,&&\text{and} &&\pid{2}=\frac{2 \chi \dho^2}{\zh} \sum_{n=1}^\infty n^2\,\bar{\gamma}_n\,.\label{phipp}
\end{align}

\subsection{Horizon constraints}
In order to make contact with the regularity condition (\ref{ConRicciFinite}) in Section~\ref{Sect:DistanceFunctionExpansion}, we first need to express the coefficients $\xi_n$ in (\ref{XiCoefs}) in terms of the $\bar{\omega}_n$. For $\rho\in[0,\dho)$ we can write
\begin{align}
\frac{1}{2\chi}\sum_{p=0}^\infty \xi_{p}\, \rho^{p}&=1+\sum_{n=1}^\infty\frac{\omega_n}{(\dho+\rho)^n}=1+\sum_{n=1}^\infty\bar{\omega}_n\,\left(\sum_{k=0}^\infty\left(-\frac{\rho}{\dho}\right)^k\right)^n\,.
\end{align}
In order to extract the term of order $\rho^p$ on the right-hand side of this equation, we use relation (\ref{PowerExpansionPower}), such that
\begin{align}
&\xi_0=2\chi\left(1+\sum_{n=1}^\infty\bar{\omega}_n\right)=\zh\,,&&\xi_p=\frac{2\chi}{(-\dho)^p p!}\sum_{n=1}^\infty\bar{\omega}_n\,\frac{(n+p-1)!}{(n-1)!}\hspace{0.5cm}\forall p\geq 1\,.\label{XiRelOmega}
\end{align}
A similar analysis allows to express the coefficients $\theta_p$ in eq.~(\ref{ThetaCoefs}) in terms of the $\bar{\gamma}_n$
\begin{align}
&\theta_0=2\chi\left(1+\sum_{n=1}^\infty\bar{\gamma}_n\right)=\zh\,,&&\theta_p=\frac{2\chi}{(-\dho)^p p!}\sum_{n=1}^\infty\bar{\gamma}_n\,\frac{(n+p-1)!}{(n-1)!}\hspace{0.5cm}\forall p\geq 1\,.\label{ThetRelGamma}
\end{align}
Furthermore, the quantity $\rootk$ in (\ref{Coefs2Extract}) takes the form
\begin{align}
\rootk=\sqrt{1-\frac{16 \chi\zh}{\dho^2}\sum_{n=1}^\infty n(n+1)\bar{\omega}_n }=\sqrt{1-\frac{16 \chi\zh}{\dho^2}\sum_{n=1}^\infty n^2\bar{\omega}_n }\,,
\end{align}
where in the last relation we have used (\ref{omega cond 1}). More generally, the  conditions (\ref{constraints}) which guarantee the finiteness of the first and second derivatives of the metric functions at the horizon, translate into
\begin{tcolorbox}[ams equation,colback=black!10!white,colframe=black!95!green]
\parbox{14cm}{${}$\\[-40pt]\begin{align}&\sum_{n=1}^\infty n\,\bar{\omega}_n=0=\sum_{n=1}^\infty n^2(n+3)\,\bar{\omega}_n\,,\hspace{1.5cm}\sum_{n=1}^\infty n\,\bar{\gamma}_n=0=\sum_{n=1}^\infty n^2(n+3)\,\bar{\gamma}_n\,,
\nonumber\\
&\hspace{2.1cm}\sum_{n=1}^\infty n^2\,\bar{\omega}_n\leq \frac{\dho^2}{16\zh\chi}\,,\hspace{1.6cm} \sum_{n=1}^\infty n^2\,\bar{\gamma}_n<\frac{(1+\rootk)\dho^2}{8\chi\zh}\,.\label{omegaconstraints}\end{align}\nonumber ${}$\\[-30pt]}
\end{tcolorbox}
In Appendix~\ref{SolvingTheSystem}, we delve into the consequences of imposing the condition that the function $f(z)$ belongs to the class $C^N(\Sigma)$, where $\Sigma$ is the submanifold defined by $z\geq \zh$ and $N\geq 2$. We also remark that the more general conditions (\ref{ConRicciFinite}) can be translated into conditions for the $\omega_n$ and $\gamma_n$, using the identifications (\ref{XiRelOmega}) and (\ref{ThetRelGamma}). Furthermore,  Appendix~\ref{SolvingTheSystem} provides a minimal solution for the system of equations obtained by truncating the series in (\ref{Functionsfh}) after $N$ terms while imposing regularity of $N$ derivatives of the metric functions at the horizon. It also discusses the limit $N\to \infty$.

\subsection{Thermodynamics}
\label{Sect:Thermo}
Next we consider the Hawking temperature \eqref{Tempkappa}: using expressions \eqref{fpH} and eliminating $\zh$ through the relations \eqref{omega zh} we have
\begin{align}
T_\mathrm{H}=&\frac{1}{8 \sqrt{2} \pi \chi \left(1+ \sum_{n=1}^\infty \bar{\omega}_n\right)}\times \nonumber \\ &  \times \sqrt{1-\frac{16 \chi ^2}{\dho^2} \left(1+\sum _{n=1}^\infty \bar{\gamma} _n\right) \left(\sum _{n=1}^\infty n^2\bar{\gamma} _n\right)+\sqrt{1-\frac{32 \chi ^2}{\dho^2} \left(1+\sum _{n=1}^\infty \bar{\omega} _n\right)\left( \sum _{n=1}^\infty n^2\bar{\omega} _n\right)}}\  ,\label{HawkingTempOms}
\end{align}
which is still complicated to evaluate directly. However, we can gain more intuition into this temperature by expanding for large mass $\chi\gg 1$. In this case, we expect that the distance of the horizon compared to the classical case is only modified by subleading terms, \emph{i.e.} $\dho=\pi\chi+\mathfrak{o}(\chi)$, which we assume to also hold true for the position of the horizon itself: $z_H=2\chi+\mathfrak{o}(\chi)$. With relation (\ref{omega zh}) this implies
\begin{equation}
2\chi\sum_{n=1}^\infty \bar{\omega}_n=\mathfrak{o}(\chi)=2\chi\sum_{n=1}^\infty \bar{\gamma}_n\ , \qq{\emph{i.e.}} \lim_{\chi\to \infty}\sum_{n=1}^\infty\bar{\omega}_n=0=\lim_{\chi\to \infty}\sum_{n=1}^\infty\bar{\gamma}_n\ .
\end{equation}
In the following, we shall furthermore assume that this is due to a genuine scaling property of the coefficients $\bar{\omega}_n$ and $\bar{\gamma}_n$ such that
\begin{equation}
\lim_{\chi\to \infty}\sum_{n=1}^\infty n^r\,\bar{\omega}_n=0=\lim_{\chi\to \infty}\sum_{n=1}^\infty n^r\, \bar{\gamma}_n\ , \quad \forall r\in\mathbb{N}\ .
\end{equation} 
Under these assumptions, the leading correction to the Hawking temperature arises from the terms
\begin{equation}
T_\mathrm{H}=\frac{1}{8\pi\chi}\left[1-\frac{1}{\pi^2}\sum_{n=1}^\infty\left(4n^2(\bar{\omega}_n+\bar{\gamma}_n)+\pi^2\bar{\omega}_n\right)+\ldots\right]\ .
\end{equation}
As an example, suppose that the proper distance can be expanded for large masses as
    \begin{align}
        &\dho=\pi \chi \left( 1 + \frac{y_1}{\sqrt{\chi}}+\mathcal{O}\left( \frac{1}{\chi}\right)\right)\ ,&&\text{with} &&y_1\in\mathbb{R}\,.\label{propdistassum}
    \end{align}
We then find for particular cases
\begin{itemize}
    \item for $(\omega_1,\gamma_1)\neq (0,0)$
    \begin{equation}
       T_\mathrm{H} =  \frac{1}{8 \pi \chi}\left(1-\frac{\pi ^2 \omega _1+4 (\gamma _1+\omega _1)}{\pi ^3 \chi}+\frac{\left(\pi ^2 \omega _1  +12(\gamma_1 + \omega _1)\right)y_1}{ \pi ^3 \chi^{3/2}} + \mathcal{O}\left( \frac{1}{\chi^{2}}\right)\right)\ .
    \end{equation}
\item for $\omega_1= 0= \gamma_1$ and $(\omega_2,\gamma_2)\neq (0,0)$
    \begin{equation}
        T_\mathrm{H} =\frac{1}{8 \pi  \chi} \left(1 -\frac{\pi ^2 \omega _2 + 16(\gamma_2+ \omega _2)}{\pi ^4 \chi^2}+\frac{\left(2\pi ^2 \omega _2 +64(\gamma_2 + \omega _2) \right)y_1 }{\pi ^4 \chi^{5/2}} + \mathcal{O}\left( \frac{1}{\chi^{3}}\right)\right)\ .\label{TempOm2}
    \end{equation}
\end{itemize}
We observe that the leading correction to the Hawking temperature is determined solely by the classical term in the expansion~\eqref{propdistassum} of the proper distance, with no dependence on higher-order terms.

To compute the entropy, we resort to the first law of thermodynamics \eqref{EntropyDef}, and we consider the two cases examined above
\begin{itemize}
    \item for $(\omega_1,\gamma_1)\neq (0,0)$
    \begin{multline}
      S = 4 \pi \chi^2 \left(1+\frac{8 \gamma_1 + 2(4 + \pi^2) \omega_1}{\pi ^3 \chi}+\right.\\ \left.+\frac{2(4 \gamma_1 + (4 + \pi^2) \omega_1)^2 }{\pi ^6 \chi^2} \log( \pi^3 \chi - 4 \gamma_1 - (4 + \pi^2) \omega_1) + \mathcal{O}\left( \frac{1}{\chi^{3}}\right)\right)\ .
    \end{multline}
    \item for $\omega_1= 0= \gamma_1$ and $(\omega_2,\gamma_2)\neq (0,0)$
    \begin{equation}
        S =4 \pi \chi^2 \left(1 + \frac{(16 \gamma_2 + (16 + \pi^2) \omega_2)^2 }{\pi^4 \chi^2}\log( \pi^4 \chi^2 - 16 \gamma_2 - (16 + \pi^2) \omega_2) + \mathcal{O}\left( \frac{1}{\chi^{3}} \right)\right)\ .
    \end{equation}
We remark that this result corrects the mass expansion of the entropy provided in the previous paper \cite{Binetti:2022xdi}. There, rather than using a self-consistent approach as in the current work, an approximation of the distance function (based on the distance function of the Schwarzschild black hole) was considered. This leads to a different conclusion for the subleading corrections of the entropy (as well as the Hawking temperature (\ref{TempOm2})).

\end{itemize}

\section{Conclusions}
\label{conclusions}
In this paper, we have provided regularity conditions for generic deformations of static and spherically symmetric black hole metrics. Following \cite{Binetti:2022xdi,Bonanno:2000ep} we have considered deformations of the radially-symmetric and static Schwarzschild space-time, which are described by corrections of the metric functions as in eq.~\eqref{modifiedfh}. In order to remain invariant under the same coordinate reparametrisations as the classical geometry, it has been proposed \cite{Binetti:2022xdi,Bonanno:2000ep} that these deformations are not arbitrary functions of the radial variable $z$, but only depend on a physical distance. Focusing on the exterior of the black hole (\emph{i.e.} outside of its event horizon), we have chosen the latter to be the proper distance $\rho$ measured from the horizon.\footnote{Other choices shall be considered in upcoming work \cite{Scheme} (see also \cite{Held:2021vwd}).} Since $\rho$ is defined through the metric function $f$ itself (concretely through the differential equation (\ref{SeriesForm})), it needs to be computed in a self-consistent fashion. In this paper we have solved this problem in a region just outside of the event horizon (which we assume to be located at $\zh$ with distance $\dho$) in two different fashions
\begin{enumerate}
\item In a first approach (see Section~\ref{nhc}) we have assumed that the first and second derivative (with respect to the radial coordinate $z$) of both $f$ and $h$ at $\zh$ are finite such that both functions can be approximated by their Taylor polynomials (\ref{fexp}) and (\ref{hexp}) (which also affords a finite expansion of the distance function in eq.(\ref{rho of z})). On the one hand, since the derivative of $\rho$ with respect to $z$ is divergent at the horizon, these assumptions lead to non-trivial conditions on the deformations of the metric functions as shown in eq.~(\ref{constraints}).\footnote{We have generalised these conditions in Appendix~\ref{App:HigherOrderDerF} by assuming arbitrarily high derivatives of $f$ to remain finite at the event horizon.} On the other hand, these also ensure that important physical quantities are well behaved: indeed, the existence of the first derivative of $f$ and $h$ (at the horizon) is required for the finiteness of the Hawking temperature and the existence of the second derivatives guarantees that the Ricci-scalar (and Kretschmann scalar) are finite (such that the black hole is free of curvature singularities) at the horizon. 
\item The second approach (see Section~\ref{Sect:SeriesExpansionApproach}) is schematically summarised in Figure~\ref{Fig:SecondApproach}: we assume that (outside of the horizon of the black hole) the deformations of the metric functions allow for a series expansion in $\rho$, as in eq.~(\ref{XiCoefs}) and (\ref{ThetaCoefs}) respectively. Taking the expansion coefficients $\{\xi_n\}$ and $\{\theta_n\}$ respectively, as well as $\dho$ as input, we look for a solution of the non-linear differential equation (\ref{DiffEqSeries}) (which is equivalent to (\ref{SeriesForm})) in the form of an integer series: more precisely, writing the radial coordinate $z$ as an integer series in powers of $\rho$ (see eq.~(\ref{SeriesFormaInt})) we determine all coefficients $\{a_n\}$ recursively in terms of the coefficients $\{\xi_n\}$ and $\dho$ (see eq.~(\ref{Coefs2Extract}) and (\ref{RecursionP})). Through series reversion, this allows to write $\rho$ as the series (\ref{SeriesRho}) in half-integer powers of $(z-\zh)$, whose coefficients $\{b_n\}$ are given in (\ref{Coeffbina}). These expressions finally allow us to compute the Ricci scalar (as well as other physical quantities) in the vicinity of the black hole horizon. Finally, the consistency of this procedure, as well as the absence of a singularity of the Ricci scalar at the horizon impose the non-trivial conditions (\ref{ConRicciFinite}) on the expansion coefficients $\{\xi_n\}$ and $\{\theta_n\}$ of the deformation of the metric functions.
\end{enumerate}
\begin{center}
\begin{figure}[h!]
       \centering
\parbox{14.5cm}{
\begin{tikzpicture}
\draw[rounded corners=2ex, ultra thick] (-1.35,-2.6) rectangle (1.35,3.6);
\path[fill=cyan!50!white, rounded corners=2ex] (-1.25,2.525) rectangle (1.25,3.5);
\path[fill=green!20!white, rounded corners=2ex] (-1.25,-2.5) rectangle (1.25,2.475);
\node at (0,3) {$\dho$};
\draw[rounded corners=1ex, thick] (-1,-0.3) rectangle (1,0.3);
\node at (0,0) {$\{\xi_n\}$, $\{\theta_n\}$};
\draw[rounded corners=1ex, thick] (-1,-2.45) rectangle (1,-1.85);
\node at (0,-2.15) {$\{\omega_n\}$, $\{\gamma_n\}$};
\draw[rounded corners=1ex, thick] (-0.7,1.85) rectangle (0.7,2.4);
\node at (0,2.1) {$e^{\Phi}$, $e^{\Psi}$};
\draw[ultra thick,->] (0,1.85) -- (0,0.35);
\node[rotate=90] at (-0.3,1.1) {{\footnotesize eq.(\ref{XiCoefs})}};
\node[rotate=90] at (0.3,1.1) {{\footnotesize and (\ref{ThetaCoefs})}};
\draw[ultra thick,->] (0,-1.8) -- (0,-0.35);
\node[rotate=90] at (-0.35,-1.05) {{\footnotesize eq.(\ref{XiRelOmega})}};
\node[rotate=90] at (0.35,-1.05) {{\footnotesize and (\ref{ThetRelGamma})}};
\node at (0,-3) {{\bf input}};
\draw[ultra thick,red,->] (1.5,0.5) -- (4,0.5);
\node at (2.75,0.8) {{\footnotesize solving}};
\node at (2.75,0.2) {{\footnotesize eq.(\ref{DiffEqSeries})}};
\draw[rounded corners=2ex, ultra thick] (4.15,-1.6) rectangle (6.9,2.6);
\node at (5.525,-3) {{\bf distance function}};
\draw[rounded corners=1ex, thick] (4.4,1.3) rectangle (6.6,2.5);
\node at (5.525,2.15) {$\{a_n\}$};
\node at (5.525,1.65) {{\footnotesize eq.(\ref{Coefs2Extract}), (\ref{RecursionP})}};
\draw[ultra thick,->] (5.525,1.3) -- (5.525,-0.25);
\node[rotate=90] at (5.2,0.5) {{\footnotesize series}};
\node[rotate=90] at (5.8,0.5) {{\footnotesize reversion}};
\draw[rounded corners=1ex, thick] (4.6,-0.3) rectangle (6.4,-1.5);
\node at (5.525,-0.65) {$\{b_n\}$};
\node at (5.525,-1.2) {{\footnotesize eq.(\ref{Coeffbina})}};
\draw[ultra thick,red,->] (7.15,0.5) -- (9.65,0.5);
\draw[rounded corners=2ex, ultra thick] (9.9,-1.8) rectangle (12.6,2.8);
\node at (11.25,-3) {{\bf physical quantities}};
\path[fill=red!30!white, rounded corners=2ex] (9.95,1.45) rectangle (12.55,2.7);
\node at (11.25,2.4) {\bf Curvature};
\node at (11.25,1.85) {{\footnotesize $R$ eq.(\ref{eq: ricci scalar h e f})}};
\path[fill=red!30!yellow!80!white, rounded corners=2ex] (9.95,1.4) rectangle (12.55,-0.4);
\node at (11.25,1) {\bf Hawking};
\node at (11.25,0.6) {\bf temperature};
\node at (11.25,0) {{\footnotesize $T_{\rm H}$ eq.(\ref{Tempkappa})}};
\path[fill=red!30!blue!30!white, rounded corners=2ex] (9.95,-0.45) rectangle (12.55,-1.7);
\node at (11.25,-0.8) {\bf Entropy};
\node at (11.25,-1.3) {{\footnotesize $S$ eq.(\ref{EntropyDef})}};
\draw[ultra thick,blue,->] (11.25,2.8) -- (11.25,3.45);
\draw[ultra thick,blue,->] (11.25,2.8) -- (11.25,4.1) -- (0,4.1) -- (0,3.7);
\draw[ultra thick,blue] (5.525,2.6) -- (5.525,4.1);
\draw[ultra thick,blue,->] (5.525,2.6) -- (5.525,3.45);
\node at (5.51225,4.4) {\footnotesize {\bf conditions} eq.(\ref{ConRicciFinite})};
\end{tikzpicture}}
\caption{\emph{Schematic overview of the approach of Section~\ref{Sect:SeriesExpansionApproach} to compute the distance function $\rho$ and physical quantities of the deformed Schwarzschild geometry.}}
\label{Fig:SecondApproach}
\end{figure}
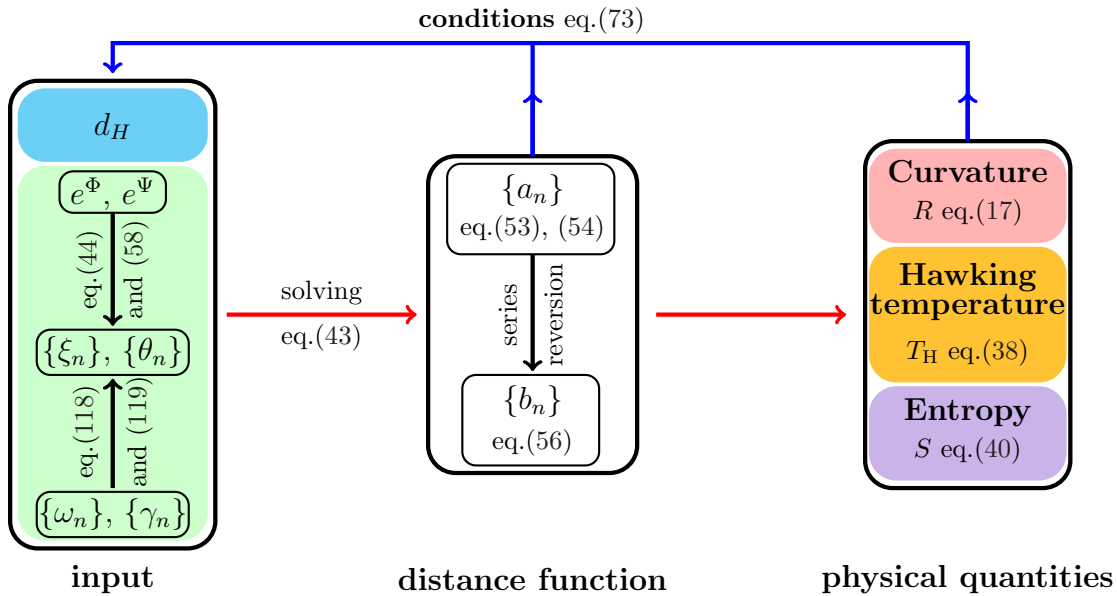
\end{center}    
\noindent
Both methods yield compatible results. They are based on assuming regularity of certain quantities (up to a given order) at the horizon and (apart from $\dho$) only require information about the black hole outside of the event horizon. They allow, however, to derive non-trivial physical quantities of the geometry (curvature scalars such as the Ricci and Kretschmann scalar) and the thermodynamics (notably the Hawking temperature and the entropy of the black hole). We have tested the conditions (\ref{constraints}) in the case of the black hole metric proposed in \cite{Bonanno:2000ep} and find that they are violated: indeed if considered as a geometry that is self-consistently defined, it leads to a divergent first derivative of the metric functions at the horizon, which poses problems for the thermodynamic interpretation, as well as a curvature singularity at the horizon. While in \cite{Bonanno:2000ep} ``approximations'' have been proposed that indeed render the geometry well defined, these correspond to a modification of the metric deformations that adhere to the conditions we have found in (\ref{ConRicciFinite}). Along these lines, other examples can be studied and we have discussed a minimal solution to these conditions.

Finally, making contact with \cite{Binetti:2022xdi}, we have considered a (generic) asymptotic expansion of the metric deformations in inverse powers of the proper distance (\ref{Functionsfh}). Assuming that the radius of convergence of the latter is sufficiently large such that these series are still valid at the event horizon, we have used the previous formalism to convert the conditions (\ref{constraints}) on the deformation functions into non-trivial relations among the asymptotic coefficients (\ref{omegaconstraints}). We have furthermore also expressed the Hawking temperature in terms of these coefficients in a consistent manner in eq.~(\ref{HawkingTempOms}), thereby correcting previous approximations in \cite{Binetti:2022xdi}.

In this paper, we have established a framework that is applicable to (quantum) deformations of the classical Schwarzschild space-time in a model-independent fashion and moreover allows to extract certain physical quantities. We have established non-trivial conditions for the deformations themselves, which can be translated into constraints in the context of concrete quantum gravity models. 

One of the most intriguing future directions is to extend our methodology to include charged and spinning black holes. Generalising our deformation approach to the classical Kerr and Reissner-Nordstr{\o}m geometries is a natural next step, which could reveal new insights into the behavior of quantum-deformed black holes with other hair parameters \cite{Carroll:2004st,Misner:1973prb}. Beyond black holes, our approach is applicable to various other space-times, such as $AdS$ spaces and even models of cosmology, for example in the context of cosmic inflation \cite{Baumann:2022mni,Vazquez:2018qdg,Albrecht:1982wi,Baumann:2008bn}. Additionally, our work opens up intriguing possibilities for studying the interior of black holes (a first hint of which is provided in Appendix \ref{App:InnerHorizon}) and the fate of the singularity at the origin within the context of quantum-deformed geometries \cite{Hawking:1976ra}.

From a broader perspective, our framework offers an exciting opportunity for quantum gravity phenomenology. By systematically extracting physical quantities and comparing them with observations, we can test and constrain concrete quantum gravity models, bridging the gap between theoretical concepts and experimental verifiability \cite{Addazi:2021xuf,Parikh:2020kfh}. Finally, it is interesting to explore the applicability of our approach beyond the realm of (quantum) gravity, for example in the context of dyons and monopoles in gauge theories.\footnote{We thank Nikita Nekrasov for suggesting this idea to us during the talk of Francesco Sannino at the First International Congress of Basic Science at the Yanqi Lake Beijing Institute of Mathematical Sciences and Applications 2023.07.16-2023.07.28.}


\section*{Acknowledgements}
We are indebted to Emanuele Binetti for collaboration at an early stage of this work and for helpful discussions and exchanges.  We furthermore thank Aaron Held for useful conversations and exchanges, and Nikita Nekrasov for relevant comments and suggestions as well as Hong-Jian He for discussions. FS wishes to thank Shing-Tung Yau for the hospitality and the organisation of the  First International Congress of Basic Science at the Yanqi Lake Beijing Institute of Mathematical Sciences and Applications 2023.07.16-2023.07.28 where this work was finalised as well as the hospitality of the CERN theoretical physics department where this work was initiated.   MDP expresses sincere appreciation to the University of Southern Denmark and D-IAS for their hospitality during the crucial final stages of the work. MDP thanks also the Galileo Galilei Institute for Theoretical Physics where part of this work was carried out. 

\appendix

\section{Regular higher order derivatives}\label{App:HigherOrderDerF}
In Section~\ref{nhc} we have derived the conditions (\ref{constraints}) by assuming that the first and second derivatives of the metric functions $f$ and $h$ are finite at the horizon $\zh$. In this appendix, we explore further conditions that stem from assuming that also higher derivatives (\emph{i.e.} beyond the second) are finite. For simplicity, we shall focus on the function $f$, while the same considerations also apply to $h$. Concretely, let $N\in\mathbb{N}$ and let us assume that all derivatives $\fn{k}$ for $k\in\{1,\ldots,N\}$ at $z=\zh$ are finite. This allows us to go beyond \eqref{fexp} and write 
\begin{align}
&f(z)=\sum_{k=1}^N \frac{\fn{k}}{k!} (z-\zh)^k+\mathcal{O}((z-\zh)^{n+1})\,, &&\text{with} &&\fn{k}:=\left.\dv[k]{f}{z}\right|_{z=\zh}\,.\label{Taylorf}
\end{align}  
\subsection{Distance function}
Inserting the expansion (\ref{Taylorf}) into the differential equation (\ref{SeriesForm}) yields a series expansion of $\rho$ in powers of $z-\zh$
\begin{align}
\rho=\sum_{k=1}^{2N-1}b_k\,(z-\zh)^{k/2}+\mathcal{O}\left((z-\zh)^{N}\right)\,,\label{TaylorPolyRho}
\end{align}
which generalises (\ref{rho of z}). The coefficients $b_k$ of this series can be found as the solutions of the differential equation (\ref{SeriesForm}) up to order $N$, which we re-write in the form
\begin{align}
\left(\dv{z}{\rho}\right)^2=\frac{1}{f(z)}\,.\label{QuadDistExpansion}
\end{align}
The right-hand side of this equation has a simple pole at $z=\zh$ and we can write the Laurent series expansion
\begin{align}
&\frac{1}{f(z)}=\sum_{m=-1}^{N-2}\ell_m\,(z-\zh)^{m}+\mathcal{O}((z-\zh)^{N-1})\,,&&\text{with} &&\ell_m\in\mathbb{R}\,.\label{InverseFexpansion}
\end{align} 
Multiplying both sides of this equation by $(z-\zh)$ and taking the limit $z\to \zh$, we find for the leading coefficient $\ell_{-1}=1/\fn{1}$. To extract the remaining coefficients, we consider the relation
\begin{align}
f(z)\,\dv{z}\frac{1}{f(z)}=-\frac{1}{f(z)}\,\dv{f}{z}
\end{align}
and expand both sides in powers of $(z-\zh)$. Comparing order by order we then find
\begin{align}
\ell_0&=-\frac{\ell_{-1}\fn{2}}{2\fn{1}}=-\frac{\fn{2}}{2(\fn{1})^2}\,,\nonumber\\
\ell_p&=\frac{-1}{(p+1)\fn{1}}\left[\frac{(p+1)\ell_{-1}\,\fn{p+2}}{(p+2)!}+\sum_{k=1}^{p-1}\frac{k\ell_k \fn{p-k+1}}{(p-k+1)!}+\sum_{k=2}^{p+1}\frac{\ell_{p-k+1}\fn{k}}{(k-1)!}\right]\,,\hspace{0.2cm}1\leq p\leq N-2\,,
\end{align}
which fixes the coefficients $\ell_p$ iteratively (in terms of the $\fn{n}$). Inserting (\ref{InverseFexpansion}) into (\ref{QuadDistExpansion}), we find the following recursive structure for the coefficients $b_p$ 
\begin{align}
&b_1=2\sqrt{\ell_{-1}}=\frac{2}{\sqrt{\fn{1}}}\hspace{2cm}b_2=0\,,\nonumber\\
&b_p=\left\{\begin{array}{lcl}\frac{2}{(2s+3)b_1}\left[\ell_p-\sum_{n=2}^{2s+2}\frac{n(2s-n+4)}{4}\,b_n\,b_{2s-n+4}\right] &\text{if}& p=2s+3\in\mathbb{N}_{\text{odd}}\,, \\ -\frac{1}{(s+1)b_1}\sum_{n=2}^{2s+1}\frac{n(2s-n+3)}{4}b_nb_{2s-n+3} &\text{if}& p=2s+2\in\mathbb{N}_{\text{odd}}\,,\end{array}\right.\hspace{0.1cm}3\leq p\leq 2N-1\,,\label{RecBF}
\end{align} 
which allows to fix them iteratively in terms of the $\fn{n}$. Since $b_2=0$, the recursive structure in (\ref{RecBF}) implies that $b_{2s}=0$ for $s\in\{1,\ldots, N-2\}$. For the odd coefficients, we find for the first few instances (for sufficiently large $N$)
\begin{align}
&b_1=\frac{2}{(\fn{1})^{1/2}}\,,&&b_3=-\frac{\fn{2}}{6(\fn{1})^{3/2}}\,,&&b_5=\frac{9(\fn{2})^2-8\fn{1}\fn{3}}{240(\fn{1})^{5/2}}\,.
\end{align}
These results are indeed compatible with (\ref{rho of z}) for the case $N=2$.

\subsection{Conditions for the regularity of higher derivatives \texorpdfstring{$f^{(n)}$}{fn}}\label{App:AllHigherOrders}
The results of the previous Subsection can be used to derive necessary conditions for the function $e^\Phi$ such that the first $N$ derivatives of the function $f$ (with respect to $z$) are finite, which is required for (\ref{Taylorf}). Assuming an expansion of the latter of the form (\ref{XiCoefs}), we have with (\ref{TaylorPolyRho})
\begin{align}
f(z)=1-\frac{1}{z}\sum_{n=0}^{2N}\xi_n\,\left(\sum_{k=1}^{2N-1}b_k\,(z-\zh)^{k/2}\right)^n+\mathcal{O}\left((z-\zh)^N\right)\,,\label{fFormExpansion}
\end{align}
which is only a function of $z$. Due to the fact that this expression contains half-integer powers of $(z-\zh)$, derivatives of $f$ can contain negative powers unless certain conditions for the coefficients $\xi_n$ are satisfied. To understand these conditions, we first re-write the summation in (\ref{fFormExpansion}), taking into account that $b_{2s}=0$ for $s\in\{1,\ldots,N-2\}$
\begin{align}
f(z)=1&-\frac{1}{z}\sum_{m=0}^N\xi_{2m}(z-\zh)^m\left(\sum_{r=0}^{N-1}b_{2r+1}\,(z-\zh)^{r}\right)^{2m}\nonumber\\
&-\frac{1}{z}\sum_{m=1}^{N}\xi_{2m-1}\left(\sum_{r=0}^{N-1}b_{2r-1}\,(z-\zh)^{r+1/2}\right)^{2m-1}+\mathcal{O}\left((z-\zh)^N\right)\,.\label{fFunSplit}
\end{align}
With the expansion of $1/z$
\begin{align}
\frac{1}{z}=\frac{1}{\zh}\sum_{n=0}^N\frac{(-1)^n}{\zh^n}\,(z-\zh)^n+\mathcal{O}\left((z-\zh)^{N+1}\right)\,,
\end{align}
it is clear that the terms in the first line of (\ref{fFunSplit}) contain no half-integer powers of $(z-\zh)$ and thus cannot contribute to singularities of derivatives of $f$ at $z=\zh$. The terms in the second line of (\ref{fFunSplit}), however, contain terms that lead to negative powers of $(z-\zh)$ for derivatives of $f$. The conditions to eliminate these singular terms for all $\fn{k}$ for $k\in\{1,\ldots,N\}$ are therefore
\begin{align}
\xi_{2n-1}=0\hspace{1cm} \forall n\in \{1,\ldots,N\}\,.\label{XiHigherRel}
\end{align}
To see this, let $s\in\{1,\ldots,N\}$ and assume that $\xi_{2s-1}\neq 0$, while $\xi_{2m-1}=0$ $\forall m\in\{1,\ldots,s-1\}$. In this case, $\fn{s}$ has a singularity of the form $(z-\zh)^{-1/2}$
\begin{align}
\dv[s]{z}f(z)&=-\frac{\xi_{2s-1}}{\zh}\,b_1^{2s-1}\,\dv[s]{z}(z-\zh)^{s-1/2}+\mathcal{O}((z-\zh)^0)\nonumber\\
&=-\frac{(2s-1)!!}{2^s\zh}\,\xi_{2s-1}\,b_1^{2s-1}\,(z-\zh)^{-1/2}+\mathcal{O}((z-\zh)^0)\,.
\end{align}
Absence of this singular contribution therefore requires $\xi_{2s-1}=0$.

Using (\ref{XiRelPhi}), the conditions (\ref{XiHigherRel}) can also be rewritten as conditions for the derivatives $\fid{n}$. For completeness, we exhibit the first few such conditions (for sufficiently large $N$)
\begin{align}
&\fid{1}=0\,,&&\fid{3}=-6\,\dho\,\fid{2}\,,&&\fid{5}=480 \dho^3 \fid{2}-20 \dho \fid{4}\,,
\end{align}
which are indeed compatible with (\ref{constraints}).


\section{Towards a fully regular solution}\label{SolvingTheSystem}
In this Appendix, we consider a particular solution for the consistency conditions found at the end of the previous Appendix. We consider a metric function $f(z)$ of the form \eqref{modifiedfh}, with deformation $e^{\Phi(z)}$, which is $N$-times differentiable (with $N\in\mathbb{N}$) in the region $z\geq \zh$, \emph{i.e.} outside of the horizon of the black hole. This allows us to define the Taylor polynomial of $f$ for large $z$
\begin{equation}\label{fz trunc N}
 f_P(z)=1-\frac{2\chi}{z}\left(1+\theta\left(\frac{1}{d(z)}\right)\right) \qq{with} \theta\left(\frac{1}{d(z)}\right)=\sum_{n=1}^N\frac{\omega_n}{d(z)^n}\ ,
\end{equation}
which we shall assume to be a satisfactory approximation of $f$ for all $z\geq \zh$. In this case, the consistency conditions at the horizon take the form of the following linear system
\begin{equation}
\sum_{n=1}^N\bar{\omega}_n=\frac{\zh(N)-2\chi}{2\chi}\qq{and} \sum_{n=1}^N\frac{(n+2k)!}{(n-1)!}\,\bar{\omega}_n=0\ ,\quad \forall k\in\{0,\ldots,N-2\}\ .\label{FiniteSystem}
\end{equation}
where we have used the rescaled coefficients $\bar{\omega}_n$ defined in \eqref{CoeffsRescale}. Furthermore, $\zh(N)$ is the position of the (external) event horizon computed from the Taylor polynomial \eqref{fz trunc N}, which is therefore implicitly a function of $N$.

\begin{figure}[htbp]
\begin{center}
\includegraphics[width=7.5cm]{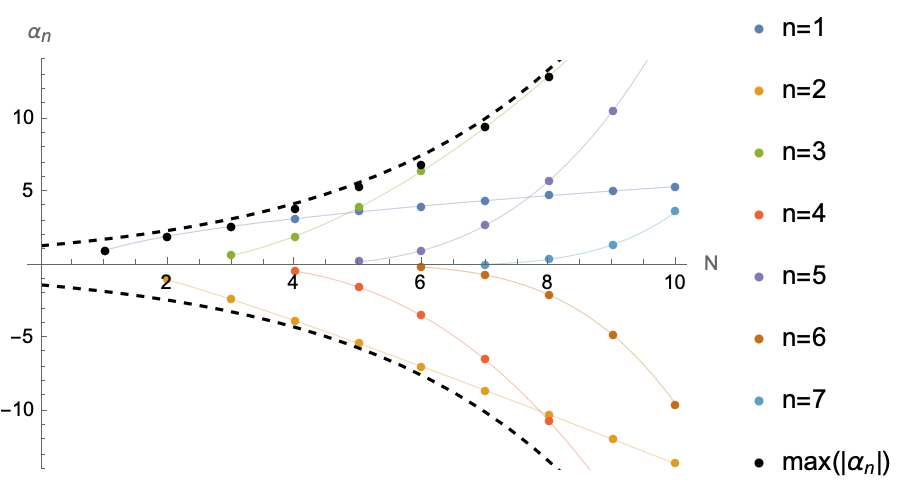}\hspace{1cm}\includegraphics[width=7.5cm]{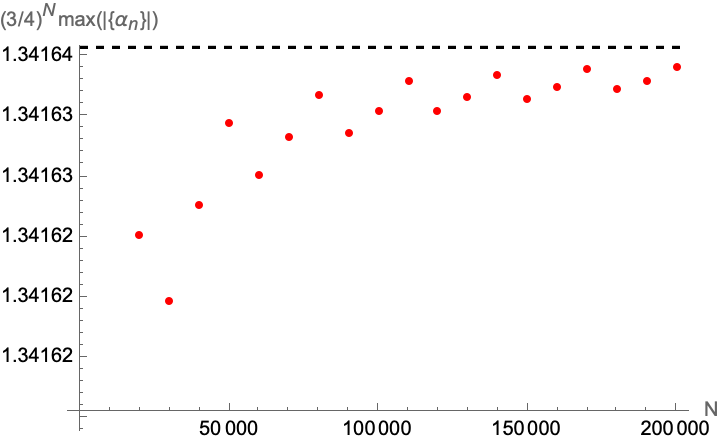}
\end{center}
\caption{\emph{Left panel: coefficients $\alpha_n$ for low values of $n$. The black dots represent $\text{max}(|\alpha_1|,\ldots,|\alpha_N|)$, while the black dashed curve represents the enveloping value according to eq.~\eqref{MaxCoeffBehaviour}. Right panel: evaluation of $|\alpha_{n_{\text{max}}}|$ in eq.~\eqref{MaxCoeff}. The dashed line represents the value $3/\sqrt{5}$.}}
\label{Fig:CoeffsLargeN}
\end{figure}

For fixed $N$, the system of equations \eqref{FiniteSystem} uniquely fixes the coefficients $\bar{\omega}_n$, for $n=1,  \ldots, N$, and we have found empirically
\begin{equation}
    \bar{\omega}_n=\frac{\zh(N)-2\chi}{2\chi}\,\alpha_n(N)\ ,
\end{equation}
with
\begin{equation}
\hspace{1cm}\alpha_n(N)=\sum_{s=0}^{n-1}(-1)^s\begin{pmatrix} n-1\\ s \end{pmatrix}\frac{\sqrt{\pi}\,\Gamma(N+\tfrac{s}{2})}{\Gamma(N-\tfrac{1}{2})\Gamma(\tfrac{s}{2}+1)}\ , \quad \forall n\in\{1,\ldots,N\}\ ,
\end{equation}
which we have tested up to $N=250$. The individual coefficients $\alpha_n$ are plotted for low values of $n$ in the left panel of Fig.~\ref{Fig:CoeffsLargeN}.

For $N\to \infty$ individual coefficients are divergent, for example, the leading contribution behaves as
\begin{equation}
\alpha_n \sim (-1)^{n-1}\sqrt{\pi}\,\frac{N^{n/2}}{\Gamma\left(\frac{n+1}{2}\right)} \qq{for} n\ll N\ .
\end{equation}
For given $N$, the largest coefficient (in terms of its absolute value) appears to be
\begin{equation}
\text{max}(|\alpha_1|,\ldots,|\alpha_N|)=\left|\alpha_{n_{\text{max}}}\right| \qq{with} n_{\text{max}}=\left\lfloor\frac{N-1}{3}\right\rfloor+1\ .\label{MaxCoeff}
\end{equation}

\begin{wrapfigure}{l}{0.48\textwidth}
${}$\\
\parbox{7.5cm}{\includegraphics[width=7.5cm]{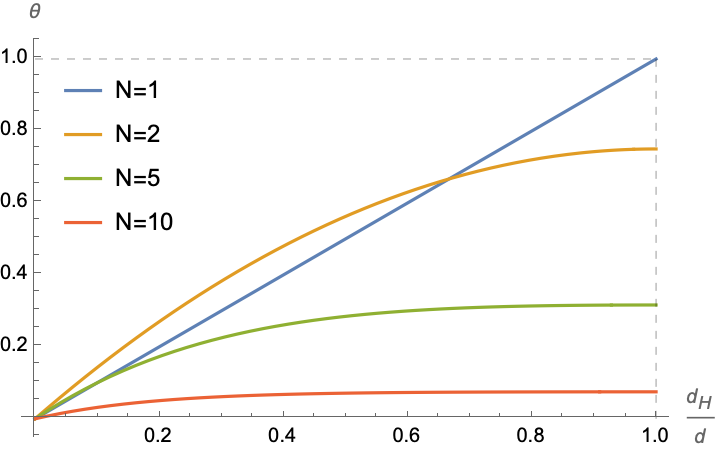}
\caption{\emph{Function $\theta$ in eq.~(\ref{fz trunc N}) for different $N$ and for the choice $\delta_0=4/3$.}}
\label{Fig:ThetaFunction}}
${}$\\[-1cm]
\end{wrapfigure}

\noindent
Numerical evaluations (see right panel of Fig.~\ref{Fig:CoeffsLargeN}) suggest that this coefficient behaves as
\begin{equation}
\left|\alpha_{n_{\text{max}}}\right|\sim \alpha_0\,\left(\frac{4}{3}\right)^N \qq{with} \alpha_0\approx \frac{3}{\sqrt{5}}\ .\label{MaxCoeffBehaviour}
\end{equation}

Therefore, for \eqref{FiniteSystem} to make sense also for large $N$, \emph{i.e.} to yield finite $\bar{\omega}_n$, the difference $\zh(N)-2\chi$ needs to tend to zero as well. Indeed, divergent coefficients $\bar{\omega}_n$ would indicate the non-existence of the derivatives of $f$ up to order $N$ and therefore contradict our initial assumptions. For example, using \eqref{MaxCoeffBehaviour}, in order for all the coefficients $\bar{\omega}_n$ (for $n\in\{1,\ldots,N\}$) to remain finite (and not tend to $0$), we may choose the following asymptotic form for the difference of $\zh$ and the classical position of the horizon
\begin{equation}
\frac{\zh(N)-2\chi}{2\chi}\sim \delta_0\,\left(\frac{3}{4}\right)^N\ ,\label{ScalingPosition}
\end{equation} 
for some constant $\delta_0\in\mathbb{R}$. The results are shown in Figure~\ref{Fig:ThetaFunction}. Indeed, for large $N$, \emph{i.e.} for a metric function that is infinitely differentiable for all $d\geq\dho$, the solution simply approaches $\theta\to 0$, \emph{i.e.} simply leads to the Schwarzschild black hole.

Notice, that through \eqref{ScalingPosition}, the condition of an infinitely differentiable metric function outside of the horizon, imposes a non-trivial condition for the position of the horizon. Through
\begin{equation}
\dho=\int_0^{\zh}\frac{\dd z}{\sqrt{|f(z)|}} \qq{and} f(\zh)=0\ ,
\end{equation}
the latter in principle enters also into relations containing the metric function inside the black hole horizon.

\section{Series relations}\label{App:SerExpansions}
For the reader's convenience, we compile several series identities in this Appendix, which are too lengthy to be presented in the main body of the paper.
\subsection{Series expansion of inverse radial coordinate}
In this Appendix, starting from the series (\ref{SeriesFormaInt}), we express the inverse radial coordinate as a series expansion in $\rho$. More precisely, we determine the coefficients $\zi{m}$ in
\begin{align}
P(\rho):=\frac{1}{z(\rho)} =\sum_{m=0}^\infty \zi{m}\,\rho^m\,,\, &&\text{with} &&\begin{array}{l}\zi{m}\in\mathbb{R}\hspace{0.5cm}\forall m\in\mathbb{N}\,,\\ \zi{0}=1/\zh\,.\end{array}\label{SeriesFormaInvZ}
\end{align}
Differentiating both sides with respect to $\rho$ and multiplying by $z$, we find
\begin{align}
z(\rho)\,P(\rho)'&=-P(\rho)\,\dv{z}{\rho}\,,
\end{align}
which with (\ref{SeriesFormaInt}) becomes the following series identity
\begin{align}
\left(\zh+\sum_{n=2}^\infty a_n\,\rho^n\right)\left(\sum_{m=1}^\infty m\,\zi{m}\,\rho^{m-1}\right)=-\left(\sum_{m=0}^\infty\zi{m}\,\rho^m\right)\left(\sum_{n=2}^\infty n\, a_n\,\rho^{n-1}\right)\,.
\end{align}
Re-arranging both sides of this equation, we find
\begin{align}
0=\zh\,\zi{1}+2(\zh\,\zi{2}+a_2\,\zi{0})\,\rho+\sum_{p=2}^\infty(p+1)\,\left[\zh\,\,\zi{p+1}+\zi{0}\,a_{p+1}+\sum_{m=1}^{p-1}\zi{m}\,a_{p+1-m}\right]\,\rho^p\,.
\end{align}
Order by order we therefore obtain the relations
\begin{align}
&\zi{1}=0\,,&&\zi{2}=-\frac{a_2}{\zh^2}\,,&&\zi{p}=-\frac{a_p}{\zh^2}-\frac{1}{\zh}\sum_{m=1}^{p-2}\,\zi{m}\,a_{p-m}\hspace{0.5cm}\forall p\geq 3\,.\label{1zRecursive}
\end{align}

\subsection{Power of a power series}
For $\rho\in[0,\dho)$ and $n\in\mathbb{N}$, we consider the following power series
\begin{align}
\left(\sum_{k=0}^\infty\left(-\frac{\rho}{\dho}\right)^k\right)^n=\sum_{p=0}^\infty \mathfrak{c}_p\,\rho^p\,.
\end{align} 
For the coefficients $\mathfrak{c}_n$ we find the following explicit expression
\begin{align}
\mathfrak{c}_p:=\left(\sum_{k=0}^\infty\left(-\frac{\rho}{\dho}\right)^k\right)^n\bigg|_{\rho^p}=\left(-\frac{1}{\dho}\right)^p\,\left(\begin{array}{c}n+p-1\\p\end{array}\right)=\left(-\frac{1}{\dho}\right)^p\frac{(n+p-1)!}{p!(n-1)!}\,.\label{PowerExpansionPower}
\end{align}
For $p=0$, we indeed have $\mathfrak{c}_0=1$. In order to show (\ref{PowerExpansionPower}) for $p \in\mathbb{N}$, we begin by demonstrating
\begin{align}
\left(\sum_{k=0}^{p-\ell}\left(-\frac{\rho}{\dho}\right)^k\right)^\ell\bigg|_{\rho^{p-\ell}}=\left(-\frac{1}{\dho}\right)^{p-\ell}\,\frac{(p-1)!}{(p-\ell)!(\ell-1)!}\,,&& \begin{array}{l}\forall p\in\mathbb{N}\,,\\\forall \ell\in\{0,\ldots,p\}\,.\end{array}\label{LemmaExpansion}
\end{align}
For $p\geq 1$ it can be verified for all (finitely) many values of $\ell\in\{0,\ldots,p\}$, concretely:
\begin{itemize}
\item for $\ell=0$ both sides are  vanishing, since we use the convention $1/((-1)!)\to 0$.
\item for $\ell=1$ we find directly
\begin{align}
\sum_{k=0}^{p}\left(-\frac{\rho}{\dho}\right)^k\bigg|_{\rho^{p-1}}=\left(-\frac{1}{\dho}\right)^{p-1}\,, 
\end{align}
which indeed agrees with (\ref{LemmaExpansion}).
\item for general $2\leq \ell\leq p$ we have
\begin{align}
&\left(\sum_{k=0}^{p-\ell}\left(-\frac{\rho}{\dho}\right)^k\right)^\ell\bigg|_{\rho^{p-\ell}}=\sum_{u_{2,\ldots,\ell}=0}^{p-\ell} \left(\sum_{k_1=0}^{p-\ell-u_2-\ldots-u_\ell}\left(-\frac{\rho}{\dho}\right)^{k_1}\bigg|_{\rho^{p-\ell-u_2-\ldots-u_\ell}}\right)\nonumber\\
&\hspace{0.2cm}\times \left(\sum_{k_2=0}^{u_2}\left(-\frac{\rho}{\dho}\right)^{k_2}\bigg|_{\rho^{u_2}}\right)\times\ldots\times  \left(\sum_{k_\ell=0}^{u_\ell}\left(-\frac{\rho}{\dho}\right)^{k_\ell}\bigg|_{\rho^{u_\ell}}\right)=\left(-\frac{1}{\dho}\right)^{p-\ell}\,\sum_{{u_{2,\ldots,\ell}=0}\atop{u_2+\ldots+u_\ell\leq p-\ell}}^{p-\ell}1\,.\nonumber
\end{align}
The last summation can be re-written in the form
\begin{align}
\sum_{{u_{2,\ldots,\ell}=0}\atop{u_2+\ldots+u_\ell\leq p-\ell}}^{p-\ell}1=\sum_{k_1=0}^{p-\ell}\sum_{{u_{3,\ldots,\ell}=0}\atop{u_3+\ldots+u_\ell\leq k_1}}1=\sum_{k_1=0}^{p-\ell}\sum_{k_2=0}^{k_1}\ldots\sum_{k_{\ell-1}=0}^{k_{\ell-2}}1=\sum_{k_1=0}^{p-\ell}\sum_{k_2=0}^{k_1}\ldots\sum_{k_{\ell-2}=0}^{k_{\ell-3}}\left(\begin{array}{c}k_{\ell-2}+1 \\ k_{\ell-2}\end{array}\right)\,,\nonumber
\end{align}
which, upon using the identity
\begin{align}
\sum_{u=0}^k\left(\begin{array}{c}u+s \\ u\end{array}\right)=\left(\begin{array}{c}k+s+1 \\ k\end{array}\right)&&\forall s\in\mathbb{N}\,,
\end{align}
leads to 
\begin{align}
\sum_{{u_{2,\ldots,\ell}=0}\atop{u_2+\ldots+u_\ell\leq p-\ell}}^{p-\ell}1=\left(\begin{array}{c}p-1 \\ p-\ell\end{array}\right)=\frac{(p-1)!}{(p-\ell)!(\ell-1)!}\,.
\end{align}
This result therefore indeed demonstrates (\ref{LemmaExpansion}).
\end{itemize}
With the result (\ref{LemmaExpansion}) we can prove (\ref{PowerExpansionPower}) (for $p\geq 1$). To this end, we consider
\begin{align}
&\left(\sum_{k=0}^\infty\left(-\frac{\rho}{\dho}\right)^k\right)^n\bigg|_{\rho^p}=\left(\sum_{k=0}^p\left(-\frac{\rho}{\dho}\right)^k\right)^n\bigg|_{\rho^p}=\sum_{\ell=0}^n\left(\begin{array}{c}n \\ \ell\end{array}\right)\left(\sum_{k=1}^p\left(-\frac{\rho}{\dho}\right)^k\right)^\ell\bigg|_{\rho^p}\nonumber\\
&\hspace{2cm}=\sum_{\ell=0}^n\left(-\frac{\rho}{\dho}\right)^{\ell}\left(\begin{array}{c}n \\ \ell\end{array}\right)\left(\sum_{k=0}^{p-1}\left(-\frac{\rho}{\dho}\right)^k\right)^\ell\bigg|_{\rho^p}\nonumber\\
&\hspace{2cm}=\sum_{\ell=1}^n\left(-\frac{\rho}{\dho}\right)^{\ell}\left(\begin{array}{c}n \\ \ell\end{array}\right)\left[\left(\sum_{k=0}^{p-\ell}\left(-\frac{\rho}{\dho}\right)^k\right)^\ell\bigg|_{\rho^{p-\ell}}\right]\,.
\end{align}
With the relation (\ref{LemmaExpansion}) we therefore obtain
\begin{align}
\left(\sum_{k=0}^\infty\left(-\frac{\rho}{\dho}\right)^k\right)^n\bigg|_{\rho^p}&=\left(-\frac{1}{\dho}\right)^p\sum_{\ell=1}^{\text{min}(n,p)}\left(\begin{array}{c}n\\ \ell \end{array}\right)\,\frac{(p-1)!}{(p-\ell)!(\ell-1)!}\nonumber\\
&=\left(-\frac{1}{\dho}\right)^p\frac{1}{p}\sum_{\ell=1}^{\text{min}(n,p)}\ell\,\left(\begin{array}{c}n \\ \ell\end{array}\right)\left(\begin{array}{c}p \\ \ell\end{array}\right)=\left(-\frac{1}{\dho}\right)^p\,\left(\begin{array}{c}n+p-1 \\ p\end{array}\right)\,,
\end{align}
which indeed demonstrates (\ref{PowerExpansionPower}).

\subsection{Series expansions for \texorpdfstring{$\xi_1\neq 0$}{xi1}}\label{App:SerNonZeroXi}
For completeness, we shall generalise the approach of Section~\ref{Sect:DistanceFunctionExpansion} also to accommodate functions $e^\Phi$ such that the coefficient $\xi_1\neq 0$ in eq.~(\ref{XiCoefs}) is non-zero. Indeed, assuming that $z\sim \zh+\mathcal{O}(\rho^p)$ for some $p\in\mathbb{R}_+$, for the left-hand-side of the equation (\ref{DiffEqSeries}) to have a term of order $\mathcal{O}(\rho)$ requires either $p=1$ or $p=3/2$. Since we have seen in Section~\ref{Sect:DistanceFunctionExpansion} that $p=1$ still requires $\xi_1=0$, we shall here explore the case $p=3/2$, \emph{i.e.} instead of (\ref{SeriesFormaInt}) we shall consider the expansion
\begin{align}
&z=\zh+\sum_{n=1}^{\infty}\ha{n}\,\rho^{\frac{n+2}{2}}\,,&&\text{with} &&\ha{n}\in \mathbb{R}\hspace{0.2cm}\forall n\in\mathbb{N}\,,\label{SeriesOdd}
\end{align}
which as before we assume to have an interval of convergence $\rho\in[0,\rho_A)$ (with some $\rho_A>0$). We then obtain
\begin{align}
z\left(1-\left(\dv{z}{\rho}\right)^2\right)&=\zh-\frac{9}{4}\,\rho\,\zh\,\ha{1}^2+\rho^{3/2}\left(\ha{1}-6\,\zh\,\ha{1}\,\ha{2}\right)+\rho^2\,\left[\ha{2}-\zh\left(4\,\ha{2}^2+\frac{15}{2}\,\ha{1}\,\ha{3}\right)\right]\nonumber\\
&\hspace{0.5cm}+\sum_{p=5}^\infty\rho^{p/2}\bigg[\ha{p-2}-\zh\sum_{m=1}^{p-1}\frac{(m+2)(p-m+2)}{4}\,\ha{m}\,\ha{p-m}\nonumber\\
&\hspace{2.6cm}-\sum_{k=1}^{p-4}\ha{k}\sum_{m=1}^{p-k-3}\frac{(m+2)(p-k-m)}{4}\,\ha{m}\,\ha{p-k-m-2}\bigg]\,.\label{SerExpOddPower}
\end{align}
Comparing the coefficients of $\rho^0$, $\rho^{1}$, $\rho^{3/2}$ and $\rho^2$ to (\ref{XiCoefs}) yields the relations
\begin{align}
&\zh=\xi_0\,,&&\xi_1=-\frac{9}{4}\,\zh\,\ha{1}^2\,,&&0=\ha{1}-6\,\zh\,\ha{1}\,\ha{2}\,,&&\xi_2=\ha{2}-\zh\left(4\,\ha{2}^2+\frac{15}{2}\,\ha{1}\,\ha{3}\right)\,,
\end{align}
which has solution
\begin{align}
&\ha{1}=\pm \frac{2}{3}\sqrt{-\frac{\xi_1}{\zh}}\,,&&\ha{2}=\frac{1}{6\zh}\,,&&\ha{3}=\frac{1-18\,\zh\,\xi_2}{135\,\ha{1}\,\zh^2}\,.\label{LowSolutionsHa}
\end{align}
The coefficient $\ha{1}$ is real only for $\xi_1<0$. In the case $\xi_1>0$, there exists no real solution of (\ref{DiffEqSeries}), which is of the form (\ref{SeriesOdd}). In the following, we shall assume $\xi_1<0$ and furthermore pick the positive sign for $\ha{1}$ in (\ref{LowSolutionsHa}): indeed, for the negative sign, the function $z(\rho)$ would not be monotonically growing for $\rho>0$.

Comparing the remaining terms in (\ref{SerExpOddPower}) order by order, we can express $\ha{p-1}$ in terms of $\ha{k}$ with $k<p-1$
\begin{align}
\ha{p-1}&=\frac{2}{3\,\ha{1}\,\zh\,(p+1)}\bigg[\ha{p-2}-\zh\sum_{m=2}^{p-2}\frac{(m+2)(p-m+2)}{4}\,\ha{m}\,\ha{p-m}\nonumber\\
&\hspace{0.5cm}-\sum_{k=1}^{p-4}\ha{k}\sum_{m=1}^{p-k-3}\frac{(m+2)(p-k-m)}{4}\,\ha{m}\,\ha{p-k-m-2}\bigg]-\left\{\begin{array}{lcl} 0 & \text{if} & p\in\mathbb{N}_{\text{odd}} \\ \frac{2\xi_{p/2}}{3\ha{1}\,\zh\,(p+1)} & \text{if} & p\in\mathbb{N}_{\text{even}} \end{array}\right.\label{SolutionsHa}
\end{align}

\subsection{Examples}
To showcase the approach developed in Section~\ref{Sect:DistanceFunctionExpansion}, we consider three simple examples, corresponding to different choices of the coefficients $\xi_n$.
\subsubsection{Schwarzschild distance}
\begin{figure}[htbp]
\begin{center}
\includegraphics[width=8cm]{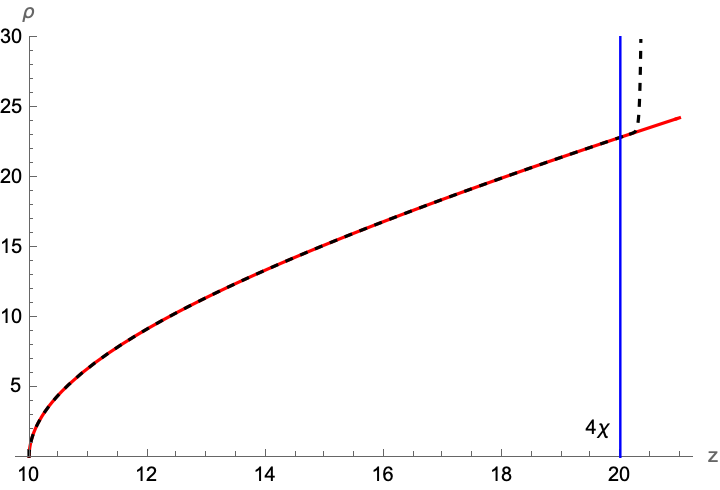}
\end{center}
\caption{\emph{Comparison of the series expansion (\ref{SeriesRho}) up to order $\mathcal{O}((z-\zh)^{500-1/2})$ (dashed black line) to the analytic result (\ref{AnalyticRhoSchwarzschild}) (red curve) for $\chi=5$. The blue line $4\chi=20$ denotes the boundary of the interval of convergence for the series expansion.}}
\label{Fig:SchwarzschildExDist}
\end{figure}

The simplest choice is to set $\xi_0=2\chi$ and $\xi_n=0$ $\forall n>0$, which corresponds to the Schwarzschild black hole (\emph{i.e.} $e^{\Phi}=1$). In this case, we also choose $\zh=2\chi$ and $\dho=\pi\chi$. Using (\ref{Coefs2Extract}) and (\ref{RecursionP}), the first few coefficients $a_{2n}$ (and their reversions $b_n$) can be tabulated as follows 
\begin{center}
\begin{tabular}{|c||c|c|c|c|c|c|}\hline
  & $n=1$ & $n=2$  & $n=3$ & $n=4$ & $n=5$ & $n=6$ \\ \hline\hline
 &&&&&&\\[-12pt]
 $a_{2n}$ & $\tfrac{1}{8\chi}$ & $-\tfrac{1}{384\chi^3}$ & $\tfrac{11}{92160\chi^5}$ & $-\tfrac{73}{10321920 \chi^7}$ & $\tfrac{887}{1857945600\chi^9}$ & $-\tfrac{136883}{3923981107200\chi^{11}}$ \\[6pt]\hline
 &&&&&&\\[-12pt]
 $b_{2n-1}$ & $2\sqrt{2\chi}$ & $\tfrac{1}{3\sqrt{2\chi}}$ & $-\tfrac{1}{40\chi \sqrt{2\chi}}$ & $\tfrac{1}{224\chi^2 \sqrt{2\chi}}$& $-\tfrac{5}{4608\chi^3 \sqrt{2\chi}}$ & $\tfrac{7}{22528\chi^4 \sqrt{2\chi}}$ \\[6pt]\hline
\end{tabular}
\end{center}
which allows to compute a series expansion of the distance function, as in eq.~(\ref{SeriesRho}). These coefficients have previously been obtained in \cite{Meissel,Foong_2008} (see also \cite{Peetre1997ErnstMA}) in a different context, namely solutions for free-falling bodies in Newtonian gravity. We have verified up to $(z-2\chi)^{500-1/2}$ that these coefficients follow the pattern
\begin{align}
b_{2n-1}=\frac{(-1)^n}{\sqrt{2\chi}}\,\frac{(2n-5)!!}{4^{n-2}\chi^{n-2}(2n-1)(n-1)!}\,,
\end{align} 
such that the series expansion of $\rho$ in terms of $(z-\zh)^{1/2}$ has an interval of convergence of $z\in [0,4\chi]$. A graphical example (for $\chi=5$) is shown in Figure~\ref{Fig:SchwarzschildExDist}, with an expansion up to order $(z-\zh)^{500-1/2}$. We have furthermore verified that the coefficients $b_{n}$ agree with a series expansion of $\rho(z)$: indeed, in the case of the Schwarzschild geometry, the  distance $\rho$ can in fact be computed in closed form as a function of $z$ (see eq.~(\ref{SchwarzschildDistance}))
\begin{align}
&\rho(z)=\sqrt{z(z-2\chi)}+2\chi\, \text{arctanh}\sqrt{1-\frac{2\chi}{z}} \qq{for} z>2\chi\,,\label{AnalyticRhoSchwarzschild}
\end{align}
allowing us to verify whether (\ref{SeriesRho}) is indeed a good representation of the distance function (see Figure~\ref{Fig:SchwarzschildExDist}).

\subsubsection{Bonanno-Reuter black hole}\label{Sect:BRDistanceComputations}
\begin{figure}[htbp]
\begin{center}
\includegraphics[width=8cm]{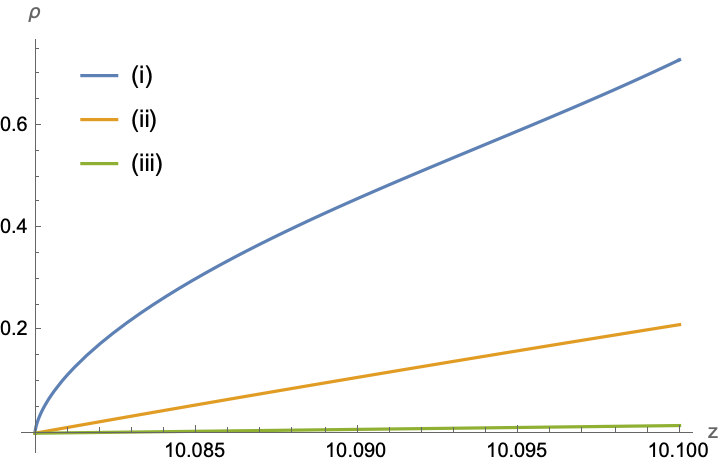}
\end{center}
\caption{\emph{Comparison of three different distance functions for the Bonanno-Reuter space-time: \emph{(i)} $\rho_{\text{BR}}$ (represented by the expansion (\ref{BRSeriesExpansion})), \emph{(ii)} the Schwarzschild proper distance $\rho_{\text{S}}=d_{\text{S}}-\dho$ and \emph{(iii)} the approximating function $\kappa_{\text{BR}}-\dho$ in eq.~(\ref{BRdistanceApprox}). Here we have chosen $\widetilde{\omega}=-1$, $\chi=5$ and $\zh=10.08$ and $\dho=12.96$.}}
\label{Fig:BRdistance}
\end{figure}

For the choice (\ref{BRmetricfunction}) of the metric deformation (along with $f=h$), the coefficients $\xi_n$ in the expansion (\ref{XiCoefs}) are given by
\begin{align}
\xi_0&=\frac{2\chi\,\dho^2}{\widetilde{\omega}+\dho^2}\,,\nonumber\\
\xi_n&=\frac{(-1)^n\,2\chi}{(\widetilde{\omega}+\dho^2)^{n+1}}\,\sum_{k=1}^{\lfloor\frac{n}{2}+1\rfloor}(-1)^k\,\left(\begin{array}{c}n+1 \\ 2k-1\end{array}\right)\,\dho^{n+2-2k}\,\widetilde{\omega}^k\,,&&\forall n\geq 1\,.
\end{align}
Here the interval of convergence (for $\widetilde{\omega}<0$) is given by $\rho\in\left[0,\frac{|\dho^2+\widetilde{\omega}|}{\dho+\sqrt{-\widetilde{\omega}}}\right)$. Moreover, since $\xi_1=\frac{4\chi \widetilde{\omega}\,\dho}{(\widetilde{\omega}+\dho^2)^2}\neq 0$, the results of Section~\ref{Sect:DistanceFunctionExpansion} cannot be directly applied. However, as explained in Appendix~\ref{App:SerNonZeroXi}, for $\widetilde{\omega}<0$ (such that $\xi_1<0$), this approach can be adapted, leading to a series expansion (\ref{SeriesOdd}) with coefficients $\ha{n}$ in (\ref{LowSolutionsHa}) and (\ref{SolutionsHa}). This expansion provides a solution of the differential equation (\ref{DiffEqSeries}) (albeit with a divergent first derivative $\fn{1}$ of $f$ at the horizon). Inversion of the series (\ref{SeriesOdd}) leads to an expansion of the proper distance to the horizon of the form
\begin{align}
\rho_{\text{BR}}(z)=\sum_{n=0}^\infty\,\widehat{b}_n\,(z-\zh)^{\frac{2+n}{3}}\,,\label{BRSeriesExpansion}
\end{align}
where for concreteness, we provide explicitly the first few coefficients
\begin{align}
&\widehat{b}_0=\frac{1}{\widehat{a}_1^{2/3}}=\frac{3^{2/3} \zh^{1/3}(\dho+\widetilde{\omega})^{2/3}}{2^{4/3}(-\dho \chi \widetilde{\omega})^{1/3}}\,,&&\widehat{b}_1=\frac{2\widehat{a}_2}{3\widehat{a}_1^2}=\frac{(\dho+\widetilde{\omega})^2}{16\dho \chi \widetilde{\omega}}\,,&&\widehat{b}_2=\frac{7\widehat{a}_2^2-6\widehat{a}_1\widehat{a}_3}{9 \widehat{a}_1^{10/3}}\,.
\end{align}
A numerical plot of the three different distance functions used in Section~\ref{Sect:ExampleBonannoReuter} for the Bonanno-Reuter black hole is shown in Figure~\ref{Fig:BRdistance}.

\subsubsection{Minimal example}
As a last example, we consider the metric function characterised by the minimal choice (\ref{Phifuns}), leading to the function $\Phi$ in eq.~(\ref{FormExpPhiExample}). In order to make contact with the approach in Section~\ref{Sect:SeriesExpansionApproach}, we first remark that the coefficients $\kappa_n$ in the series expansion of (\ref{FormExpPhiExample}) (see (\ref{SerPhiKappa})) are explicitly given by
\begin{align}
&\kappa_n=-\frac{(-1)^n\,(n-1)(n-3)}{2\dho^{n+2}}\,\phi_2\,,&&\forall \,n\geq 0\,.\label{MinExCoefsKappa}
\end{align}
Therefore, the series expansion (\ref{SerPhiKappa}) of (\ref{FormExpPhiExample}) has interval of convergence $\rho\in[0,\dho)$. Using (\ref{XiRelPhi}), the coefficients $\kappa_n$ allow to calculate the expansion coefficients $\xi_n$ of $e^{\Phi}$. The first few $\xi_n$ read explicitly
\begin{align}
&\xi_0=2\chi\,e^{-\frac{3\phi_2}{2\dho^2}}\,,&&\xi_1=0\,,&&\xi_2=\frac{2\chi}{\dho^4}\,\phi_2\,e^{-\frac{3\phi_2}{2\dho^2}}\,,&&\xi_3=0\,, \label{CoefsMinModel}
\end{align}
which therefore satisfy the conditions (\ref{ConditionsfhRegular}), as expected. The leading coefficients $a_n$ in the expansion of $z(\rho)$ in (\ref{SeriesFormaInt}) therefore become
\begin{align}
&a_2=\frac{1+\rootk}{8\zh}\,,&&a_3=-\frac{\zh \phi_2}{\dho^4(1+3\rootk)}\,,&&a_4=-\frac{1}{1+2\rootk}\left(\frac{(1+\rootk)^3}{256\zh^3}+\frac{9\zh^3 \phi_2^2}{\dho^8(1+3\rootk)^2}\right)\,,
\end{align}
where we have again used the shorthand notation $\rootk=\sqrt{1-16\zh\xi_2}$. The series inversion yields the following coefficients for the distance function (\ref{SeriesRho})
\begin{align}
&b_1=\frac{2\sqrt{2\zh}}{\sqrt{1+\rootk}}\,,&&b_2=\frac{32\zh^3\phi_2}{\dho^4(1+\rootk)^2(1+3\rootk)}\,,&b_3=\frac{\dho^8(1+\rootk)^4(1+3\rootk)^3+256\zh^6(19+29\rootk)\phi_2^2}{2\sqrt{2\zh}\dho^8(1+\rootk)^{7/2}(1+2\rootk)(1+3\rootk)^2}\,.
\end{align}

\section{Further examples}\label{Sect:FurtherExamples}
In this Appendix, we discuss two further examples from the literature that describe non-singular, static, and spherically symmetric black holes: the first is the Hayward black hole \cite{Hayward_2006} and the second one is the Dymnikova space-time \cite{Dymnikova:1992ux}. Although we are aware that these examples do not exhibit any divergent physical quantities, it is still interesting to demonstrate how our approach can be applied to deformation functions that explicitly depend on coordinates other than the proper distance. 

\subsection{Hayward black hole}
We begin by examining the Hayward black hole, which was introduced in \cite{Hayward_2006} as the first model to describe a non-singular black hole (notably at the origin) without committing to any specific modification of General Relativity. The metric function for the Hayward space-time can be written as
\begin{align}
&f(z)=h(z)=f_{\text{Hay}}(z):=1-\frac{2\chi z^2}{z^3+2\chi \gamma} \, ,&&\forall z\in[0,\infty)\,,
\end{align}
where $\gamma$ is a free parameter that determines the scale at which the departure from the classical Schwarzschild solution becomes significant \cite{Frolov:2016pav}. Here we assume $f_{\text{Hay}}$ to hold in the entire space-time.
From the metric element, we can directly deduce the form of the deformation function, which explicitly depends on the coordinate $z$ rather than the proper distance $d$
\begin{equation}
    e^{\Phi\left(\frac{1}{d(z)} \right)} := \frac{z^3}{z^3+2\chi \gamma}\ .
\end{equation}
The Hayward space-time exhibits two event horizons, indicated by the existence of two zeroes of the function $f_\mathrm{Hay}$. The position of the outer horizon is:
\begin{align}
\zhp=\frac{2\chi}{3}\,\left(1+2\cos\left[\frac{1}{3}\,\arccos\left(1-\frac{27\gamma}{8\chi^2}\right)\right]\right)\,.
\end{align}
Furthermore, the proper distance can be computed using the integral expression \eqref{propd}:
\begin{equation}\label{HaywardPropDist}
    d_\mathrm{Hay}(z)= \int_0^z \sqrt{\abs{\frac{\tilde{z}^3+2\chi \gamma}{ \tilde{z}^3 - 2 \chi (\tilde{z}^2 + \gamma)} }}\dd \tilde{z} \ .
\end{equation}

\begin{figure}[!h]
    \centering
    \includegraphics[width=0.75\textwidth]{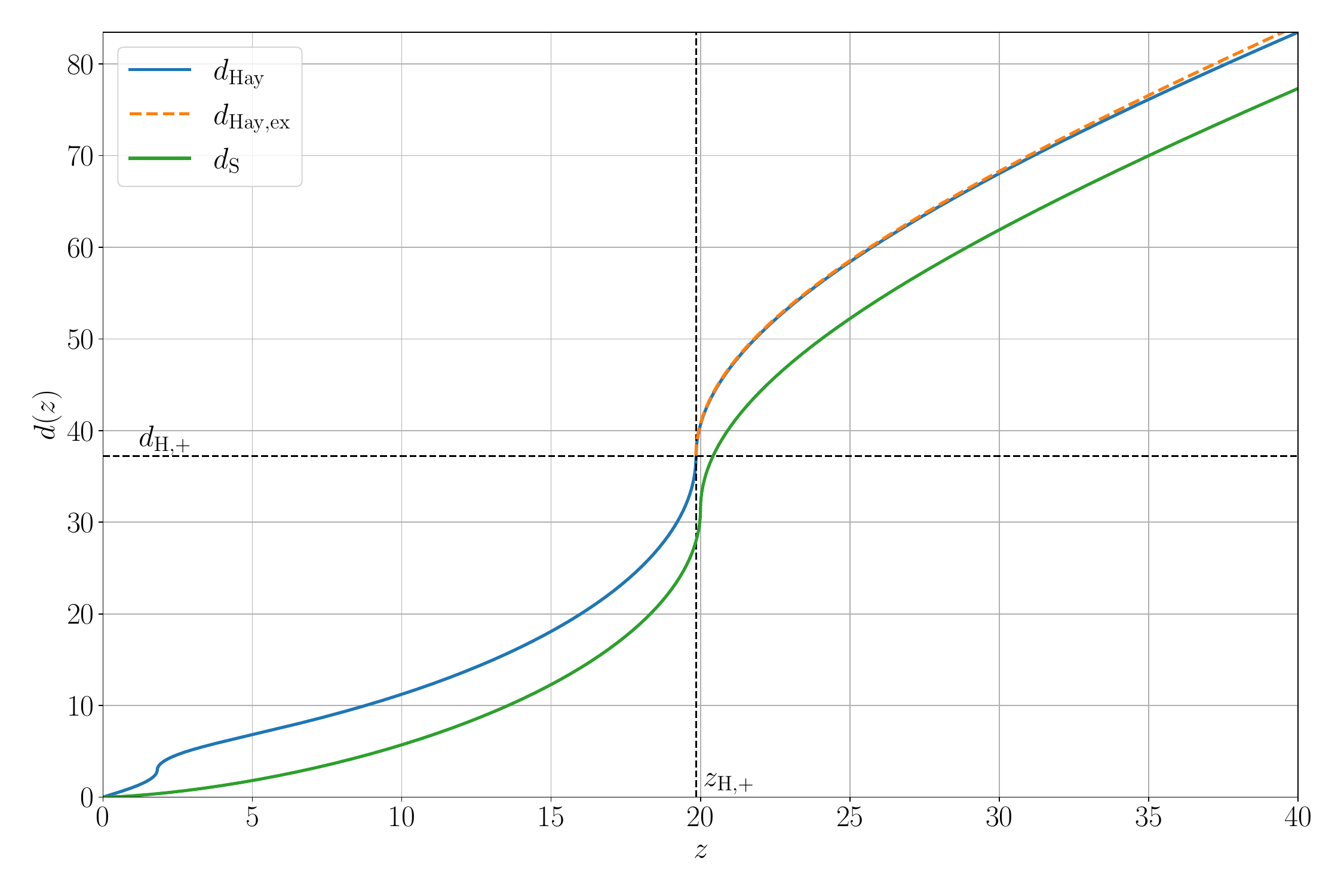}
    \caption{\emph{Comparison between the proper distance in the Hayward space-time (blue, calculated using equation \eqref{HaywardPropDist}) and the proper distance in the Schwarzschild space-time $d_\mathrm{S}$ (green) for specific parameter values: $\chi=10$ and $\gamma=3$. The dashed orange line corresponds to the Taylor series expansion of the proper distance from the event horizon, obtained from equation \eqref{DistHayExpansion}.}}
    \label{HaywardDistance}
\end{figure}

Numerical evaluation of this integral suggests the following form of the distance of the horizon for large values of $\chi$
\begin{align}
&\dhp=\pi \chi+c_1\, \gamma^{c_2}\,\ln \chi+\mathcal{O}(\chi^0)\,,&&\text{with} &&\begin{array}{l}c_1=0.3334\pm 0.0002\,, \\ c_2=0.4999\pm 0.0002\,.\end{array}
\end{align}
More importantly, the distance (for a generic point outside of the horizon) can be expanded in powers of $(z-\zhp)$, which takes the explicit form
\begin{align}
d_{\text{Hay,ex}}&=\dhp+\frac{\sqrt{2}(\zhp^3+2\gamma\chi) \sqrt{z-\zhp}}{\sqrt{\zhp \chi(\zhp^3-4\gamma\chi)}}+\frac{\chi(\zhp^6-14\zhp^3\gamma\chi+4\gamma^2\chi^2)}{3\sqrt{2}(\zhp\chi(\zhp^3-4\gamma\chi))^{3/2}}\,(z-\zhp)^{3/2}\nonumber\\
&\hspace{1.35cm}+\mathcal{O}((z-\zhp)^{5/2})\,.\label{DistHayExpansion}
\end{align}
A comparison between the numerical solution for the distance function \eqref{HaywardPropDist} and \eqref{DistHayExpansion} is shown in Figure~\ref{HaywardDistance}.
Using expansion \eqref{SeriesRho} we can read off the coefficients $b_n$ from eq.~\eqref{DistHayExpansion} and express them in terms of the $a_n$'s using the recursive relation listed in~\eqref{Coeffbina}
\begin{equation}
   a_2^{-1/2}=\frac{\sqrt{2}(\zhp^3+2\gamma\chi)}{\sqrt{\zhp \chi(\zhp^3-4\gamma\chi)}}\, , \,  - \frac{a_3}{2 a_2^2} = 0\, , \, \frac{5 a_3^2 - 4 a_2 a_4}{8 a_2^{7/2}} = \frac{\chi(\zhp^6-14\zhp^3\gamma\chi+4\gamma^2\chi^2)}{3\sqrt{2}(\zhp\chi(\zhp^3-4\gamma\chi))^{3/2}}\, .
\end{equation}
Solving this system yields the solutions for $a_2$, $a_3$ and $a_4$ 
\begin{align}
    a_2 &= \frac{\zhp \chi (\zhp^3 - 4 \gamma  \chi)}{2(\zhp^3 + 2 \gamma \chi)^2} \, , \qquad a_3 = 0 \, , \\ a_4 &= \frac{-\chi ^2 \zhp^{10}+18 \gamma  \chi ^3 \zhp^7-60 \gamma ^2 \chi ^4 \zhp^4+16 \gamma ^3 \chi ^5 \zhp}{384 \gamma ^5 \chi ^5+12 \zhp^{15}+120 \gamma  \chi  \zhp^{12}+480 \gamma ^2 \chi ^2 \zhp^9+960 \gamma ^3 \chi ^3 \zhp^6+960 \gamma ^4 \chi ^4 \zhp^3}\ .
\end{align}
Plugging these coefficients into the series \eqref{SeriesRho} allows us to write $z$ as a power series in $\rho$ and then expand around $\rho=0$
\begin{multline}
\frac{2\chi z(\rho)^3}{z(\rho)^3+2\chi \gamma} = \zhp + \frac{6 \gamma  \chi ^3 \zhp^3 \left(\zhp^3-4 \gamma  \chi \right)}{\left(2 \gamma  \chi +\zhp^3\right)^4} \rho^2 +\\+ \frac{\gamma  \chi ^4 \left(-7 \zhp^{12}+72 \gamma  \chi  \zhp^9-204 \gamma ^2 \chi ^2 \zhp^6+112 \gamma ^3 \chi ^3 \zhp^3\right)}{\left(2 \gamma  \chi +\zhp^3\right)^7}\rho^4 + \mathcal{O}(\rho^5)\ ,
\end{multline}
from which we can read the coefficients $\xi_n$. In particular, we obtain 
\begin{equation}
  \xi_1=0\, , \quad \xi_2 = \frac{6 \gamma  \chi ^3 \zhp^3 \left(\zhp^3-4 \gamma  \chi \right)}{\left(2 \gamma  \chi +\zhp^3\right)^4} \qq{and} \xi_3 = 0\ .
\end{equation}
Alternatively, we can use the last relation in equation \eqref{InitialConditionsGenericExpansion} and the recursive relation in equation \eqref{RecursionP} for $p=3$ to determine the values of the coefficients $\xi_2$ and $\xi_3$. By doing so, we can establish that the Hayward space-time satisfies the condition \eqref{ConditionsfhRegular}. Consequently, it is not surprising that the Ricci scalar and the Hawking temperature in this space-time are well-defined and free from singularities at the event horizon $\zhp$.
\subsection{Dymnikova black hole}
We now turn our attention to the Dymnikova space-time, which was proposed in \cite{Dymnikova:1992ux,Dymnikova:2004qg}. This space-time describes a static, spherically symmetric non-singular black hole embedded in an effective energy-momentum tensor. 
The metric function in the Dymnikova space-time is given by
\begin{align}
f(z)=h(z)=f_{\text{Dymn}}(z):=1-\frac{2\chi}{z}\,\left(1-e^{-z^3/z_*^3}\right) \qq{with} \begin{array}{l}z_\ast^3 = 2 \chi z_0^2\, ,\\ \forall z\geq 0\,,\end{array}
\end{align}
which (as for the Hayward black hole) we take to hold for the entire space-time. Also, similar to the Hayward black hole, the parameter $z_0$ ensures the regularity of the solution near the origin. Additionally, the density profile of the effective energy-momentum tensor is chosen such that the metric possesses a de Sitter core.
Furthermore, the Dymnikova space-time possesses two horizons, which for large masses are located at
\begin{equation}
    \zhp = 2\chi\left(1 - \mathcal{O}\left( e^{-4 \chi^2/z_0^2} \right) \right) \qq{and} \zhm = 2\chi \left(1 - \mathcal{O} \left(\frac{z_0}{8 \chi} \right) \right)\ .
\end{equation}

\begin{figure}[!h]
    \centering
    \includegraphics[width=0.75\textwidth]{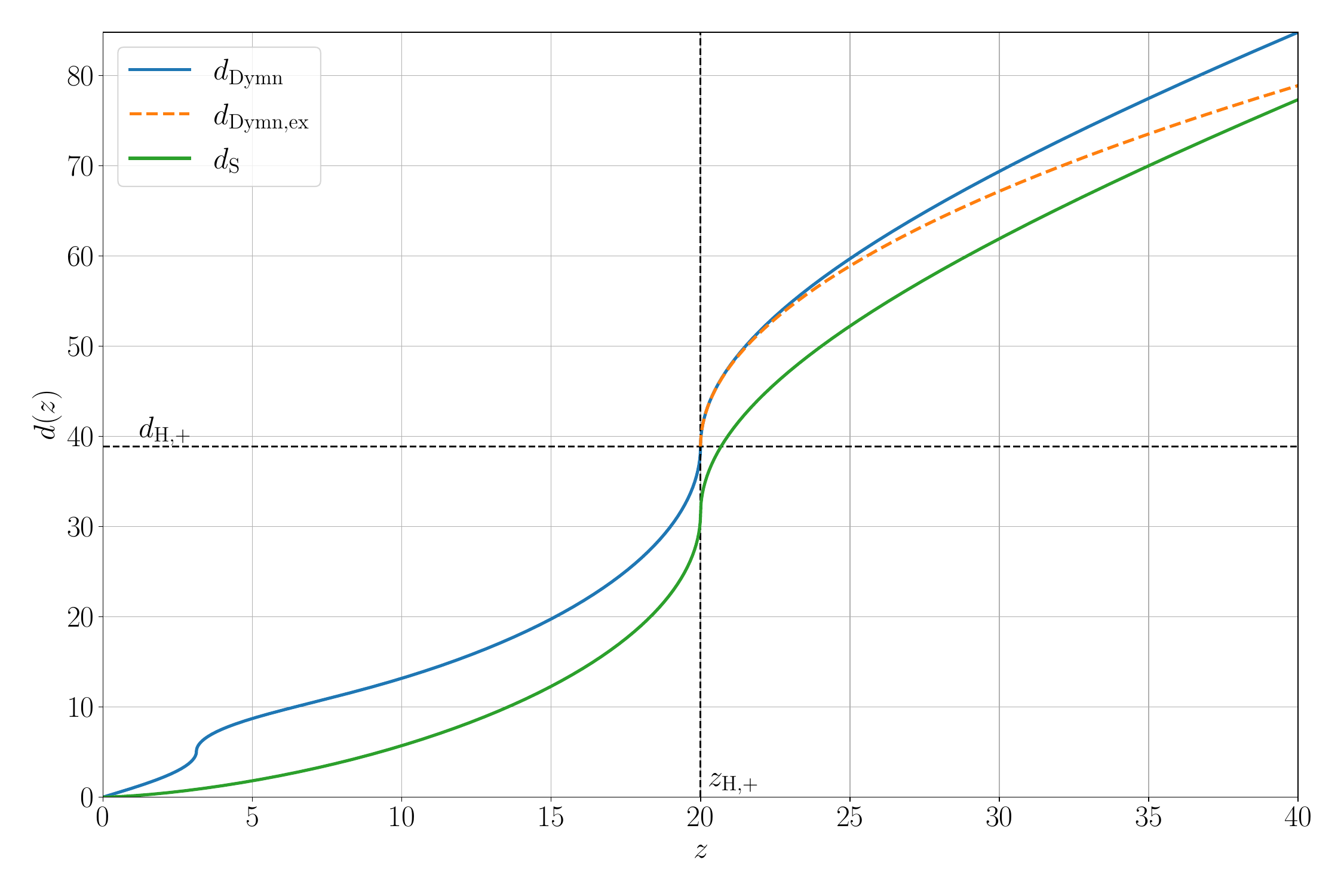}
    \caption{\emph{Comparison of the proper distance in the Dymnikova space-time, computed using the definition \eqref{propd} (blue), with the proper distance in the Schwarzschild space-time $d_\mathrm{S}$ (green) for $\chi=10$ and $z_0=3$. The dashed orange line represents the Taylor series expansion of the proper distance from the event horizon, as given by equation \eqref{DistDymnExpansion}.}}
    \label{figDymnikova}
\end{figure}

We find from \eqref{rho of z} that the power series expansion of the proper distance has the following form
\begin{multline}
    d_\mathrm{Dymn,ex} = \dhp + \frac{2 \zhp \sqrt{z-\zhp}}{\sqrt{2 \chi - e^{-\zhp^3/2 \chi  z_0^2} \left(2 \chi + 3 \zhp^3/z_0^2 \right)}}  + \\ + \frac{e^{-\zhp^3/ 2 \chi  z_0^2} \left(8 \chi ^2 z_0^4 \left(e^{ \zhp^3/ 2 \chi  z_0^2}-1 \right)-9 \zhp^6\right)}{24 \chi  z_0^4 \left(2 \chi +e^{- \zhp^3/ 2 \chi  z_0^2} \left(-2 \chi - {3 \zhp^3}/{z_0^2}\right)\right)^{3/2}}(z-\zhp)^{3/2} + \mathcal{O}((z-\zhp)^2)\ .\label{DistDymnExpansion}
\end{multline}
In Figure \ref{figDymnikova} we provide a numerical approximation of the proper distance using the definition \eqref{propd} and compare this expansion for $z > \zhp$. Furthermore, we can recover the coefficients $a_2$, $a_3$ and $a_4$
\begin{align}
    &a_2 = \frac{e^{-\zhp^3/2 \chi  z_0^2} \left(2 \chi  z_0^2 \left(e^{\zhp^3/2 \chi  z_0^2}-1\right)-3 \zhp^3\right)}{4 z_0^2 \zhp^2}\, , \qquad a_3=0 \, ,\\
    & a_4 = \frac{e^{-\zhp^3/2 \chi  z_0^2} \left(2 \chi  z_0^2 \left(e^{\zhp^3/2 \chi  z_0^2}-1\right)-3 \zhp^3\right) \left(9 \zhp^6-8 \chi ^2 z_0^4 \left(e^{\zhp^3/2 \chi  z_0^2}-1\right)\right)}{384 \chi  z_0^6 \zhp^5}\ .
\end{align}
As for the Hayward space-time, we can expand the function $2\chi(1- e^{-z(\rho)^3/z_\ast^3})$ around $\rho=0$
\begin{equation}
    2\chi(1- e^{-z(\rho)^3/z_\ast^3}) = \zhp + \frac{3 e^{-\zhp^3/ \chi  z_0^2 } \left(2 \chi  z_0^2 \left(e^{\zhp^3/ 2 \chi  z_0^2 }-1\right)-3 \zhp^3\right)}{4 z_0^4} \rho^2 + \mathcal{O}(\rho^4)
\end{equation}
from which we read off the coefficients $\xi_1$, $\xi_2$ and $\xi_3$
\begin{equation}
    \xi_1 = 0\ , \quad \xi_2 = \frac{3 e^{-\zhp^3/ \chi  z_0^2 } \left(2 \chi  z_0^2 \left(e^{\zhp^3/ 2 \chi  z_0^2 }-1\right)-3 \zhp^3\right)}{4 z_0^4} \qq{and} \xi_3 = 0\ .
\end{equation}
We observe that the conditions stated in equation \eqref{ConditionsfhRegular} are satisfied in the Dymnikova space-time as well, leading to the same conclusions as those drawn for the Hayward black hole. These conditions ensure that the Ricci scalar and the Hawking temperature remain well-defined and free from singularities at the event horizons, which is indeed the case for this space-time. 

\section{Conditions for regularity at inner horizons}\label{App:InnerHorizon}
The sufficient conditions (\ref{constraints}) in Section~\ref{Sect:Conditions} for regularity of the Ricci tensor at $\zh$, have been derived assuming that the latter is the position of the outer horizon: notably, we have assumed in various instances that $f(z)>0$ for $z>\zh$. Generalised Schwarzschild BHs, however, may have further horizons, which are characterised by a vanishing of the function $f(z)$ in (\ref{eq: metric}), such that the derivative of the distance (\ref{propd}) diverges. The latter can (in the same way as discussed in Section~\ref{Sect:Conditions}) lead to curvature singularities that are physically not acceptable. Assuming that the form of the metric functions $f,h$ are still of the form (\ref{modifiedfh}), the presence of an inner horizon, therefore, puts further conditions on the functions $\Phi$ and $\Psi$. We can obtain these conditions by straightforwardly generalizing the discussion of Section~\ref{Sect:Conditions}. Here we shall briefly exhibit them, assuming a black hole with two (simple) horizons at $\zhpm$ (with $\zhp>\zhm$)\footnote{Here $\zhp$ is understood to be the position of the outer horizon, which we denote by $\zh$ throughout the remainder of this paper.} and distances $\dhp$ and $\dhm$ respectively.

For simplicity, we consider the conditions that $\mathfrak{f}_1:=f'(\zhm)$ and $\mathfrak{f}_2:=f''(\zhm)$ are both finite, with $\mathfrak{f}_1<0$. Concretely, we write
\begin{align}
&f(z)=\mathfrak{f}_1\,(z-\zhm)+\frac{\mathfrak{f}_2}{2}\,(z-\zhm)^2+\mathfrak{o}\left((z-\zhm)^2\right)\,,&&\text{for} &&z>\zhm\,.
\end{align}
Furthermore, we define
\begin{align}
&d=:\dhm+\xi(z)\,,&&\text{with} &&\xi(z)=\frac{2\sqrt{z-\zhm}}{\sqrt{-\mathfrak{f}_1}}+\frac{\mathfrak{f}_2}{6}\,\frac{(z-\zhm)^{3/2}}{(-\mathfrak{f}_1)^{3/2}}+\mathfrak{o}\left((z-\zhm)^{3/2}\right)\,.\nonumber
\end{align}
For $z>\zhm$ we therefore find
\begin{align}
z=\zhm-\frac{\mathfrak{f}_1}{4}\,\xi^2+\frac{\mathfrak{f}_1\mathfrak{f}_2}{96}\,\xi^4+\mathfrak{o}(\xi^4)\,,
\end{align}
and thus we have the following conditions
\begin{align}
&\dv{\Phi}{y}\bigg|_{y=\frac{1}{\dhm}}=0\,,&&\dv[3]{\Phi}{y}+6\,\dhm\,\dv[2]{\Phi}{y}\bigg|_{y=\frac{1}{\dhm}}=0\,,
\end{align}
which can be expressed in terms of the $\Phi^{(n)}$ (where we have already taken into account (\ref{constraints}))
\begin{align}
0&=\left(7\,\dhp\,\dhm-3\dhp^2-4\dhm^2\right)\,\frac{\Phi^{(2)}}{\dhp \dhm^2}+\sum_{n=4}^\infty\frac{(\dhp-\dhm)^{n-1}}{(n-1)!\dhp^{n-1}\dhm^{n-1}}\,\Phi^{(n)}\,,\nonumber\\
0&=42(\dhm-\dhp)\Phi^{(2)}+\sum_{n=4}^\infty\frac{\left[(n+4)\dhp-6\dhm\right](\dhp-\dhm)^{n-3}}{\Gamma(n-1)\,\dhp^{n-2}\dhm^{n-3}}\,\Phi^{(n)}\,.
\end{align}


\printbibliography
\end{document}